\documentclass[5p,times,authoryear]{elsarticle}

\usepackage[spanish]{babel}     
\addto\captionsspanish{%
}
\usepackage[latin1]{inputenc}   
\usepackage{amsmath}            
\usepackage{epstopdf}           
\usepackage{flushend}           
\usepackage{graphicx}	
\usepackage{amssymb}	
\usepackage{lineno}
\usepackage{longtable}
\usepackage{mathtools}
\usepackage{paralist}
\usepackage{float}
\usepackage{euscript}
\usepackage{xcolor} 
\usepackage{soul}

\usepackage{amssymb}

\usepackage[figuresright]{rotating}
\usepackage{changes}

\usepackage{subfigure}

\begin{document}

\begin{frontmatter}






\title{Analysis of the dynamics of a spacecraft in the vicinity of an asteroid binary system with equal masses}


\author[Leo]{Santos, L. B. T.\corref{cor1}}
\ead{leonardobarbosat@hotmail.com}

\author[Pri]{Sousa-Silva, P. A.}
\ead{priscilla.silva@unesp.br}

\author[Maisa]{Terra, M. O.}
\ead{maisa@ita.br}

\author[Leo]{Aljbaae, S.}
\ead{safwan.aljbaae@gmail.com}

\author[Diogo]{Sanchez, D. M.}
\ead{dmsanchez@ou.edu}

\author[prado,prado1]{Prado, A. F. B. A.}
\ead{antonio.prado@inpe.br}

\author[geraldo]{Oliveira, G. M.}
\ead{magela@cefetmg.br}

\author[filipe]{Monteiro, F.}
\ead{filipeastro@on.br}

\author[allan]{de Almeida Jr, A. K.}
\ead{allan.junior@inpe.br}

\author[nathan,nathan1]{Lima, N.B.}
\ead{nathan.lima@ufpe.br}

\author[Nathalia,nathan1]{Lima, N.B.D.}
\ead{nathalia.blima@ufpe.br}

\cortext[cor1]{Corresponding author.}

\address[Leo]{Division of Space Mechanics and Control. National Institute for Space Research, INPE. S\~ao Paulo, Brazil}

\address[Pri]{S\~ao Paulo State University, UNESP. S\~ao Jo\~ao da Boa Vista, Brazil.}

\address[Maisa]{Technological Institute of Aeronautics, ITA. S\~ao Jos\'e dos Campos, Brazil.}

\address[Diogo]{The University of Oklahoma, Norman OK, USA.}

\address[prado]{Postgraduate Division - National Institute for Space Research (INPE), S\~ao Paulo, Brazil.}

\address[prado1]{Professor, Academy of Engineering, RUDN University, Miklukho-Maklaya street 6, Moscow, Russia, 117198.}

\address[geraldo]{Federal Center for Technological Education of Minas Gerais (CEFET-MG), Contagem, Brazil.}

\address[filipe]{National observatory, R. Gen. Jos\'e Cristino, 77 - S\~ao Crist\'ov\~ao, 20921-400, Rio de Janeiro - RJ, Brazil.}

\address[allan]{Instituto de Telecomunica\c{c}\~oes, Universidade de Aveiro, 3810-193 Aveiro, Portugal.}

\address[nathan]{Department of Physics, Federal University of Pernambuco, 50740-540, Recife, Brazil.}

\address[nathan1]{Brazilian Institute for Material Joining and Coating Technologies (INTM), Federal University of Pernambuco, 50740-540, Recife, Brazil.}

\address[Nathalia1]{Department of Fundamental Chemistry, Federal University of Pernambuco, 50740-560, Recife, Brazil.}

\begin{abstract}
In this work, we performed a dynamical analysis of a spacecraft around a nearly equal-mass binary near-Earth asteroid with application to the asteroid 2017 YE5, which is also a possible dormant Jupiter-family comet. Thus, we investigated the motion of a particle around this binary system using the circular restricted three-body problem. We calculated the locations of the Lagrangian points of the system and their Jacobi constant. Through numerical simulations, using the Poincar\'e Surface of Sections, it was possible to find several prograde and retrograde periodic orbits around each binary system's primary, some exhibiting significantly-sized higher-order behavior. We also calculated the stability of these orbits. After finding the periodic orbits, we investigated the influence of solar radiation pressure on these orbits. For this analysis, we considered that the area-to-mass ratio equals 0.01 and 0.1. 
We also performed a spacecraft lifetime analysis considering the physical and orbital characteristics of the 2017YE5 system and investigated the behavior of a spacecraft in the vicinity of this system. We analyzed direct and retrograde orbits for different values of Jacobi's constant. This study investigated orbits that survive for at least six months, not colliding or escaping the system during that time. We also analyze the initial conditions that cause the spacecraft to collide with $M_1$ or $M_2$, or escape from the system. In this work, we take into account the gravitational forces of the binary asteroid system and the solar radiation pressure (SRP).
Finally, we calculated optimal bi-impulsive orbital maneuvers between the collinear Lagrangian points. We found a family of possible orbital transfers considering times of flight between 0.1 and 1 day.
\end{abstract}


\end{frontmatter}


\section{Introduction}

Most known binary asteroids are composed of a larger primary component rotating rapidly and a small secondary component, which is normally in an almost circular orbit around the primary. This secondary component is usually referred to as a satellite. Both radar and photometric studies showed that about 15\% of the near-Earth objects (NEO) population larger than 0.3 km are binary systems \citep{2002Sci...296.1445M, 2006IcaPravecBIN, 2015aste.book..355M}. However, some binary systems are composed of two components of similar size and mass in mutual orbit about each other. To date, only four almost equal-mass near-Earth binary asteroids have been reported in the literature, including (69230) Hermes, 1994 CJ1, (190166) 2005 UP156, and 2017 YE5. These systems have a component size ratio ($D_s$/$D_p$) higher than 0.8, where the component size ratio ($D_s$/$D_p$) is the diameter of the secondary body divided by the diameter of the primary body.

Radar observations identified the NEO 2017 YE5 as a binary system during its apparition in 2018. These observations indicated that the system is composed of two components of 900 m in diameter each, separated by 1.8 km \citep{2018DPSTAYLOR, 2019LPI2945TAYLOR}. In addition, the radar data suggest that the components have slightly different shapes and some differences in surface brightness. Photometric lightcurve observations of this binary system recently 
reported in \citet{2021MNRAS.tmp.2192M} confirmed that the mutual orbital period of this system is about 24 hours, supporting radar estimates \citep{2018DPSTAYLOR, 2019LPI2945TAYLOR}. Its physical characteristics indicate that it is similar to cometary nuclei, having a dark albedo of about 2-3 percent, a featureless spectrum, classified as a D-type, and a low density from 0.6 to 1.2 g/cm$^{3}$, implying in a rubble-pile structure for the components and/or a significant presence of volatile compounds  \citep{2021MNRAS.tmp.2192M}.

The Tisserand parameter (T$_{J}$ ) is usually used to distinguish comets and asteroids \citep{1994Icar18LEVISON}. This quantity is a constant of motion in the restricted three-body problem and is computed from several orbital elements (semi-major axis, orbital eccentricity, and inclination) of a relatively small object and a larger ``perturbing body'', in this case, Jupiter. This parameter remains approximately constant during the small body's lifetime, even after encounters with planets, and is therefore useful for identifying them. Most main-belt asteroids have T$_{J}$ $>$ 3, most Jupiter-family comets (JFCs) have 2 $<$ T$_{J}$ $<$ 3, and Halley-family and long-period comets have T$_{J}$ $<$ 2 \citep{1994Icar18LEVISON}. The 2017 YE5 system has a Tisserand parameter of T$_{J}$ = 2.87, implying a comet-like orbit. According to \citet{2021MNRAS.tmp.2192M}, the 2017 YE5 system is probably a dormant/extinct Jupiter-family binary comet since its orbital and physical characteristics are similar to JFC nuclei.

Different binary NEOs were or will be targeted by spacecraft missions, including (65803) Didymos, which was the target of NASA's Double Asteroid Redirection Test (DART) mission \citep{2020Icar..34813777N}, (175706) 1996 FG3 and (35107) 1991 VH, which are the targets of the NASA's JANUS mission \citep{2021LPI....52.1706S}, and the equal-size binary Trojan (617) Patroclus, which is one of the targets of the NASA's Lucy mission \citep{2021PSJ.....2..171L}. Most of these binaries are primitive asteroids, i.e., rich in organic and volatile compounds, probably formed beyond the outer main belt, which can provide clues about the origin of water and prebiotic molecules on the early Earth \citep{2000MPS...35.1309M, 2013ApJ...767...54I}.

Several problems in astronomical systems and space dynamics can be solved using simpler models, such as the Circular Restricted Three-Body Problem (CRTBP). Therefore, in the fields of astrophysics and astrodynamics, the CRTBP has received wide attention. The CRTBP is a nonintegrable dynamical system, and analytical solutions for the orbits are generally not available. In these cases, numerical methods are required. More accurate trajectories can be studied under more precise models (such as the polyhedron model for the gravity field of the asteroids) when considering the final stages of a mission, but a general knowledge of the gravity field based on a small number of parameters is very useful in the early stages of the project of a mission.
\cite{2023AJ....165..140L} investigated the orbital dynamics of exoplanets in the vicinity of close binary stars and applied dynamical systems theory techniques using the CRTBP. \cite{2020CSF...13409704Z} investigated the equilibrium points as well as the associated convergence basins considering the restricted problem, with modified gravitational potential. \cite{Alshaery} investigated the spatial quantized restricted three-body problem. The authors found new equilibrium points, both in the plane of the primaries and out of the plane, and investigated the zero velocity curves and their relationship with the Jacobi constant \citep{Alshaery}. One of the most important contributions of the CRTBP is the Jacobi constant that can be used to generate zero velocity curves, which allow us to know the permitted regions of the third body in space \citep{2015ApSS.357...58A}. Furthermore, \cite{babaresi} investigated quasi-periodic orbits near complex rotator asteroids. The motion of a negligible mass particle around the complex rotator 4179 Toutatis was considered as another numerical example. As a numerical example, it was considered the motion of a particle around the complex rotator 4179 Toutatis. \cite{Ahmed} carried out an analytical study of the restricted three-body problem considering the effect of the gravitational force of the asteroid belt. After developing the equation that governs the motion of a particle with negligible mass in the vicinity of the system, the authors determined the location of the equilibrium points as well as the linear stability of the motion around these points \citep{Ahmed}. 
 \cite{2020NewA...7501319A}, using the multiple scale method, found periodic solutions of the circular Sitnikov problem. Furthermore, \cite{2020NewA...7501319A} demonstrated that the initial conditions play a vital role in the approximated and numerical solutions behavior.

The Poincar\'e Surface Section (PSS) technique produced by \cite{1892mnmc.book.....P} is a widely used technique to investigate chaotic regions, as well as periodic and quasi-periodic orbits in a qualitative way. \cite{2018MNRAS.474.2452B} investigated the trajectory around the asteroid 4179 Toutatis using PSS. \cite{2010JGCD...33.1010D} have investigated periodic orbits for the Earth-Moon system using PSS considering the undisturbed case. \cite{2010Ap&SS.326..263D} investigated through numerical simulations the stability (of the $f$ family) of periodic orbits considering the Earth-Moon system using the CRTBP model using the Poincar\'e surface section as a tool.
\cite{dutt} have also been analyzing periodic orbits in the vicinity of the Lagrangian point $L_2$ using PSS. \cite{2011Ap&SS.333...37S} have been investigating periodic orbits for the Saturn-Titan system using PSS, assuming Saturn as the oblate spheroid. \cite{2020arXiv201206781A} investigated the dynamics of a spacecraft in the vicinity of the asteroid (99942) Apophis during its approach to Earth, taking into account the gravitational perturbations of the Sun, planets, and the Solar Radiation Pressure (SRP) using the PSS as a tool. The author found that, during the approach, the Earth is the one that most affect the dynamics of the space vehicle, causing the vast majority of orbits to collide or escape from the system.

\cite{2016IJAA....6..175P} and \cite{2016IJAA....6..254P} used the PSS technique to investigate periodic and quasi-periodic orbits considering the Sun-Saturn system with the actual flattening of Saturn and taking into account the SRP as perturbations. \cite{2016IJAA....6..440P} also analyzed periodic orbits around the primary bodies considering the Sun-Earth and Sun-Mars systems taking into account (as perturbations) the flattening of the secondary bodies (less massive) and the SRP, respectively, using PSS. \cite{2019JAnSc..66..475P} used PSS techniques to investigate higher-order interior resonant orbits from the perturbed photogravitational restricted three-body problem considering the Earth-Sun system. \cite{Zotos} studied the Sun-Jupiter system and used the SALI (Smaller-Alignment Index) technique in order to complement the information obtained by PSS and to classify the orbits according to their periodic/chaotic nature, as well as collisional evolution or escape through $L_1$/$L_2$.

The interest in exploring the solar system's small bodies has grown a lot in recent years. Consequently, we have seen a dramatic increase in the publications of technical-scientific articles involving space missions to asteroids and comets. When it comes to binary asteroid systems, several efforts have already been focused on the dynamics of a particle in the vicinity of this type of system.
For example, \cite{2012PSS...70..102H} investigated the stability of orbits in the vicinity of binary asteroid 175706 (1996 FG3), analyzing the complex gravity fields disturbed by solar radiation pressure (SRP). The authors modeled the gravitational field considering the fourth-degree homogeneous triaxial ellipsoidal harmonic expansion. \cite{WooP} investigated the motion of a spacecraft around a binary asteroid system. The author assumed the inertial properties of irregularly shaped bodies in order to calculate the gravitational potential of the system. \cite{2015CeMDA.122...27B} explored the dynamic structure around a binary system, which has a large mass ratio, using the CRTBP as a model. \cite{2016CeMDA.126..405X} studied the periodic forced motions produced by SRP in the vicinity of uniformly rotating asteroids. \cite{2019AcAau.163...11S} presented a numerical method to search for periodic orbits in the vicinity of the binary asteroid (66391) 1999 KW4 without taking into account the SRP. 
\cite{Vilac} classified ballistic orbits using clustering as a tool, as well as generated databases of certain orbits for small asteroid orbiters. \cite{2020MNRAS.496.1645A} investigated the prograde motion of a massless particle around the binary asteroid system (90) Antiope using numerical methods and taking into account the shape of both asteroids (polyhedral shape) and the SRP. 

Among the known binary asteroids in our Solar system, the 2017 YE5 system is unique. Observational evidence suggests that the binary 2017 YE5 is a possible extinct/dormant Jupiter family comet, being the first to be detected in this class. The binary 2017 YE5 appears to be a plausible target for a space mission, as it can provide details on the volatile and organic content in the near-Earth region and clues about differences in the formation processes of binary systems. In addition, for being a dormant comet candidate, it is an interesting target to understand the end states of comets or to investigate the dynamical processes that move asteroids from typical asteroidal orbits to cometary-like ones.

Due to the fact that this system had its physical and orbital characteristics recently discovered, and that could be an interesting target for a space mission, we investigated the orbital dynamics in the vicinity of this system using the CRTBP. Thus, the binary asteroid 2017 YE5 (NEO) was chosen for a numerical study in this paper. 

This work differs from the aforementioned literature as it investigates the motion of the spacecraft around the asteroid binary system with equal masses and considers in the analysis the physical and orbital characteristics of the asteroid binary system 2017 YE5. We developed the equations of motion using the CRTBP and calculated the positions of the equilibrium points, as seen in Sections \ref{EofM} and \ref{LP}. Furthermore, in Section \ref{PSS}, we search for periodic orbits and analyze the stability of these orbits in the vicinity of the 2017 YE5 system using the PSS as a tool. Aiming at a more general study, in Section \ref{VVT}, we investigate the variations of velocity required to open transfers between different regions considering the CRTPB. In order to make the problem more realistic, in Section \ref{SRP}, we investigate the effect of solar radiation pressure on the periodic orbits found. In addition, we build grids of initial conditions that provide information on the final destination of a spacecraft in the vicinity of the 2017 YE5 system considering different energy values. Finally, in Section \ref{transfer}, we investigate the bi-impulsive transfer between the collinear point of the 2017 YE5 system.

\section{Equations of Motion}
\label{EofM}

In this section, we define the mathematical model of the circular restricted three-body problem (CRTBP) in a rotating reference frame to analyze the dynamics close to the binary system 2017 YE5. We use this model because the bodies involved have shapes that are very close to spherical  \citep{2019LPI2945TAYLOR}.
In this model, it is assumed that the dynamics of a massless body, ($P(x,y)$), is analyzed when traveling in the gravitational field generated by an asteroid binary system, $ M_1 $ and $ M_2 $, with masses $ m_1 $, $ m_2 $, respectively.
The distances from $P$ to $M_1$ and $M_2$ are, respectively, $r_1=[(x + \mu)^2 + y^2 + z^2]^{1/2}$ and $r_2 = [(x + \mu - 1)^2 + y^2 + z^2]^{1/2}$. 

We assume that the unit of time is selected such that the orbital period of the primaries around the center of mass of the system is equal to 2$\pi$. The sum of the masses of the primaries is $ m_1 + m_2 = 1 $ in the canonical unit. Then, the mass ratio of the system is $\mu = \frac {m_2}{m_1 + m_2} = 0.5$. The positions of the primaries, $M_1$ and $M_2$, on the $x$, $y$, and $z$ axis are ($-\mu$, 0, 0) and ($1 - \mu$, 0, 0), respectively, which means (-0.5, 0, 0) and (0.5, 0, 0) in the present case of equal masses for the primaries \citep{2007IJBC...17.1151G}. Considering the rotating reference frame at the center of mass of the binary system, the angular frequency given by Kepler's law is $\omega^2R^3 = G(m_1 + m_2) \Rightarrow \boldsymbol{\omega}=(0,0,1)$. Here, $R$ is the distance between the two masses. Considering a zone where the motion of a massless particle $P$ is highly dominated by the binary's own gravitational field, that is to say, deep inside its Hill sphere, the equations of motion in the rotating reference frame are:
\begin{eqnarray} \label{eq2}
   \ddot{\textbf{r}}&=&-2\boldsymbol{\omega} \boldsymbol{\times} \dot{\textbf{r}} - \boldsymbol{\omega} \boldsymbol{\times} (\boldsymbol{\omega} \boldsymbol{\times} \textbf{r})+U_{\textbf{r}},
\end{eqnarray}
where $\textbf{r}$ is the coordinate vector of the particle, and $U_{\text{r}}$ is the gradient of the gravitational potential of the system $U$, which is given by
\begin{eqnarray} \label{eq2U}
    U = \frac{\mu}{r_1} + \frac{1-\mu}{r_2},
\end{eqnarray}
where, $r_1^2 = (x+\mu)^2 + y^2 + z^2$, $r_2^2 = (x-1+\mu)^2 + y^2 + z^2$.

The CRTBP has five equilibrium points, known as Lagrangian points, and are labeled $L_k$, $k$ = 1,2,3,4,5. The collinear points ($L_1$, $L_2$, $L_3$) are located on the $x$ axis. On the other hand, the triangular points ($L_4$, $L_5$) are located on the $xy$ plane. All Lagrangian points are located on the $z$ = 0 plane.

The CRTBP has an integral of motion known as a Jacobi integral.
\begin{eqnarray}
    C(\boldsymbol{x}, \mu) = x^2+y^2 + \frac{2\mu}{r_1} + \frac{2(1-\mu)}{r_2} - (\mbox{v}_x^2 + \mbox{v}_y^2 + \mbox{v}_z^2).
	\label{Omega}
\end{eqnarray}
If the particle's velocity becomes zero, the zero-velocity surfaces (ZVS) are defined by
\begin{eqnarray}
    C(\boldsymbol{x}, \mu) = x^2+y^2 + \frac{2\mu}{r_1} + \frac{2(1-\mu)}{r_2} = 2\Omega(x, y, z).
	\label{CVZ}
\end{eqnarray}
where $\Omega(x, y, z)$ is the modified potential.
Equation~(\ref{CVZ}) provides a mathematical expression for the ZVS in Cartesian coordinates. 
The motion of the body is only possible when $ x^2+y^2 + \frac{2\mu}{r_1} + \frac{2(1-\mu)}{r_2} > C$; otherwise, the square of a real number (the velocity) would be negative, which is not possible  \citep{SZB1967, leo, 2017MNRAS.464.3552A, 2019MNRAS.486.2557A}. For the planar motion, $z=0$, and then Eq.~(\ref{CVZ}) defines the zero-velocity curves (ZVC).

\section{Numerical results}
\label{results1}
To generate the results, we consider the motion of a spacecraft in the $xy$ plane, which rotates around the $z$ axis with an angular velocity $\omega = 1$.

\subsection{Lagrangian points}
\label{LP}
It is necessary to know the physical and orbital characteristics of the system under analysis in order to calculate the locations of the Lagrange points of a binary system.

\begin{figure}
\centering
	\includegraphics[width=0.8\columnwidth]{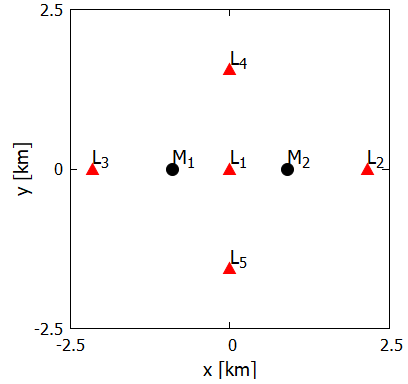}
    \caption{Image of the five Lagrange points and the locations of the primary bodies.}
    \label{ep}
\end{figure}

The location of the Lagrangian points can be found by solving the equation $\bigtriangledown \Omega(x, y, z) = 0$. Figure \ref{ep} shows, in kilometers, the distribution of the five Lagrangian points of the 2017 YE5 system (red triangles) and the location of the center of mass of each primary body (black circles). 

Table \ref{complete} shows the exact locations of the Lagrange points in the canonical unit (c.u.) and kilometers (km) and their Jacobi constant $C$ (c.u.) for the 2017 YE5 system.
				
		
		
		
		
		
	

\begin{table}
	\centering
	\caption{Locations of the Lagrange points and their Jacobi constant $C$, in the 2017 YE5 system.}
	\label{complete}
	\begin{tabular}{|c | c | c |c | c | c |}
				
		\cline{1-6}
		{ } & {$x~ (c.u.)$} &{$y ~(c.u.)$}&{$x~ (km)$}&{$y ~(km)$}&{$C~ (c.u.)$}\\
		\cline{1-6}
		
		\hline
		
		\multicolumn{1}{|c|}{$L_1$} & 0 & 0 & 0 & 0 & 4.00 \\
		\cline{1-6}
		
		{$L_2$} & 1.19841 & 0 & 2.15713 & 0 & 3.45679\\
		
		\cline{1-6}
		
		{$L_3$} & -1.19841 & 0 & -2.15713 & 0 & 3.45679\\
		
		\cline{1-6}
		
		{$L_4$} & 0 & 0.86602 & 0 & 1.55884 & 2.750\\
		
		\cline{1-6}
		
		{$L_5$} & 0 & -0.86602 & 0 & -1.55884 & 2.750\\
		\hline	
	
	\end{tabular}
 \end{table}

Each Lagrangian point is associated with a Jacobi constant. The first one ($ L_1 $) emerges when Jacobi's constant is $ C_1 = 4.00 $. This is the $C$ value that allows an orbital transfer between $M_1$ and $M_2$ under natural motion. This transfer can only take place through $ L_1 $. Note that, using $ C = 4.00 $, a transfer of a massless particle close to one of the primaries to infinity is impossible since this value of $C$ does not open the passages by the Lagrangian points $ L_2 $ and $ L_3 $. 

Reducing the value of $C$, the internal $ZVCs$ grow and the external ones decrease. When $ C $ reaches the value $ C_{2-3} = 3.45 $, the internal and external ovals of the zero velocity curves intersect each other at $ L_2 $ and $ L_3 $. Because the primaries have the same mass, these Lagrangian points arise for the same Jacobi constant ($C = 3.45$). 

A natural transfer between $ M_1 $ and $ M_2 $ is still possible and, in addition, it is also possible to transfer a spacecraft between the primaries and infinite, passing through the Lagrangian points $ L_2 $ and $ L_3 $.

By further increasing the energy of the system, i.e., decreasing the Jacobi constant, the forbidden regions become smaller, and the allowed regions of motion of the particle become larger. When $ C $ reaches the value $ C_{4-5} $ = 2.75, $ L_4 $ and $ L_5 $ appear. 
For values of $C$ less than $ C_{4-5} $, all points in the phase space are accessible.

\subsection{Periodic orbits}
\label{PSS}

The Poincar\'e Surface of Section (PSS) is plotted in order to obtain information about the solutions and overall behaviors of the dynamical system. The Figures shown in this section provide an overview of the orbital structure in the vicinity of the 2017 YE5 system. It is possible to distinguish between chaotic and regular motion.
A point on the PSS, for example, represents a periodic orbit. On the other hand, a closed curve in a specific region means the presence of a quasi-periodic orbit.
The presence of scattered isolated points suggests chaotic orbits, while areas without points indicate areas that are not accessible by any orbit. In addition, the PSS provides information to find initial conditions for periodic solutions. The PSS was determined numerically with the integrator ODE113, part of Matlab, in which it uses variable-step, optimized for the accuracy of $10^{-12}$. We stop our integration after 2500 intersections between the trajectory and the plane $y$ = 0.  

After numerically finding the PSS, we have the initial conditions (approximate) to find the periodic orbits. Using the monodromy matrix and the differential correction scheme as a tool, it is possible to find the periodic orbits as well as to analyze their stability through the characteristic multipliers related to the monodromy matrix \citep{1979CeMec..20..389B, Howell1, natasha2, 2020CeMDA.132...28Z}.


The monodromy matrix $M$ associated with periodic solutions of the CRTBP is a symplectic matrix since the CRTBP is a Hamiltonian system. 
This implies that if $\lambda \in \mathbb{C}$ is an eigenvalue of $M$ then so are $\bar{\lambda}$ and $1/\lambda$ (and hence also $1/ \bar{\lambda}$)\citep{IME_lectures}. 
Therefore the eigenvalues of $M$ are constituted by three reciprocal pairs. Indeed, since the solution is periodic and due to the existence of the first integral of motion \citep{Broucke1}, two of these six multipliers are unitary so that
\begin{equation}
\mbox{Spec} \ M=\{1, 1, \lambda_1, \lambda_1^{-1}, \lambda_2, \lambda_2^{-1} \}.    
\end{equation}

Stability indices provide a useful measure of orbital stability. Following \cite{Broucke1}, in this article we assume that the stability index is defined as $s_i$ = $|\lambda_i$ + $1/\lambda_i|$, $i$ = 1, 2. 
As well made explicit by \cite{Gomez_Mondelo_2001}, $s_j$ can be in one of the following cases:

\begin{enumerate}

\item {\sf Hyperbolic:} $\lambda_j \in \mathbb{R}$, $\lambda_j \ne \pm 1$, thus $s_j>2$.

\item {\sf Elliptic:} $\lambda_j \in \mathbb{C}$, $|\lambda_j|=1$, ($\lambda_j=e^{i \alpha}$, with  $\alpha \in \mathbb{R}$), but $\lambda_j \ne \pm 1$. 
Thus  $s_j<2$. In particular, the {\sf parabolic} case is given by $\lambda_j=\pm 1$, thus $s_j=2$. 

\item {\sf Complex unstable:} $\lambda_j \in \mathbb{C}/\ \mathbb{R}$ and $|\lambda_j|\ne 1$, such that $s_j \in \mathbb{C}/\ \mathbb{R}$. As a consequence of the structure of the symplectic matrix eigenvalues mentioned above, if $s_j$ is complex unstable, then so is $s_{3-j}$, and indeed, $s_{3-j}=\bar{s_j}$.
\end{enumerate}




A periodic orbit is stable and has no unstable subspace if $s_i$ $<$ 2, for $i=1, 2$; otherwise, the orbit is unstable \citep{Gomez_Mondelo_2001,2020CeMDA.132...28Z}.

A hyperbolic characteristic multiplier pair implies associated stable/unstable manifolds, while an elliptic pair is related to a central subspace.
We assume that $s_1$ is related to the stability indexes that are associated with the stable/unstable subspace ($\lambda^{W_s}$/$\lambda^{W_u}$). On the other hand, we assume that $s_2$ is the stability index considering the pair that represents the central subspace.

Bifurcations can occur when $s_i$ = 2 \citep{natasha2}. 
The characteristics of the types of bifurcations that appear depend on the periodic orbits' eigenvalues (characteristic multipliers) obtained from the monodromy matrix.
More details regarding characteristic multiplier collisions are beyond the scope of this article.

We will consider in this analysis the planar motion of a spacecraft. In this way, we define the PSS as the hyperplane given by 
\begin{equation}
\{(x, v_x) \ | \ y = 0 \mbox{ and } v_y>0 \mbox{ for a fixed } C \mbox{ value}\}.    
\end{equation}

%

To generate the initial condition sets, we separated the hyperplane of the PSS into three subsets as a function of $x$. For the first one, we varied $ x_0 $ from -1 to -0.55. After that, we varied $ x_0 $ from~-0.45 to 0.45. Finally, we varied $ x_0 $ from 0.55 to 1 (remembering that the primaries are in $x_1$ = -0.5 and $x_2$ = 0.5 with $y$ = 0).

\begin{figure}
	\includegraphics[width=\columnwidth]{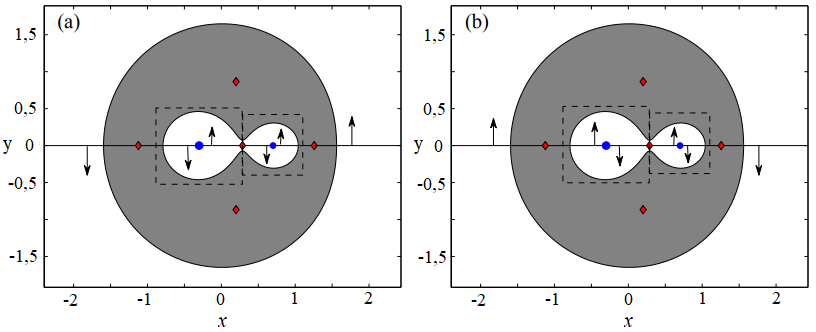}
    \caption{Definition of (a) prograde and (b) retrograde trajectories by region. Source: Taken from \citet{natasha2}.}
    \label{poinc123}
\end{figure}

Given the values of $x_0$ and null v$_{x_0}$, we determine the magnitude of v$_{y_0}$ according to the expression of the Jacobi constant given by Eq.~(\ref{Omega}). 
The choice of $v_y>0$ defines if trajectories are initially progressive or retrograde in relation to a given primary, as depicted by the black arrows in Figure \ref{poinc123}.
Between $L_1$ and $L_3$, the initial motion sense is defined with respect to $M_1$; while between $L_1$ and $L_2$, it is defined with respect to $M_2$.
Trajectories starting in the outer region are equivalently defined as prograde or retrograde with respect to any of the primaries.
As time evolves, consecutive crossings of the trajectories with the PSS are recorded and represented in the ($x$, $v_x$) plane.

In the case considered here, we take the units of mass, the distance between the primaries, and angular velocity such that $\mu$ = 0.5, $d$ = $\omega$ = 1, respectively. 
For each plot, we typically compute about 100 different initial conditions along the $x$-axis, and 3000 crossings for each initial condition.
As follows, several surfaces of section for  different $C$ Jacobi constant values are presented and studied.

\begin{figure}
	\includegraphics[width=\columnwidth]{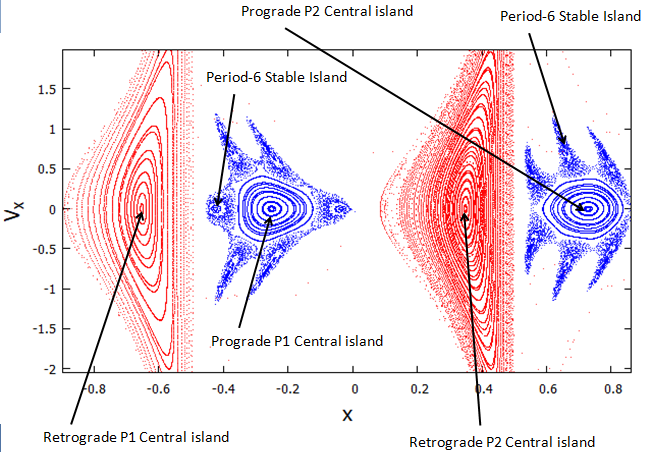}
    \caption{PSS for $C =$ 4.00 around 2017 YE5 system.}
    \label{generalC1}
\end{figure}

Figure~\ref{generalC1} shows a PSS and it represents a subset of the dynamics for $C$ = 4.  This value of the Jacobi constant is the value at the Lagrangian point $ L_1 $. 

Red points on the map correspond to trajectories that, initially, are retrograde with respect to one of the primaries. On the other hand, blue intersections show motions that are initially prograde. 

\begin{figure}
	\includegraphics[width=\columnwidth]{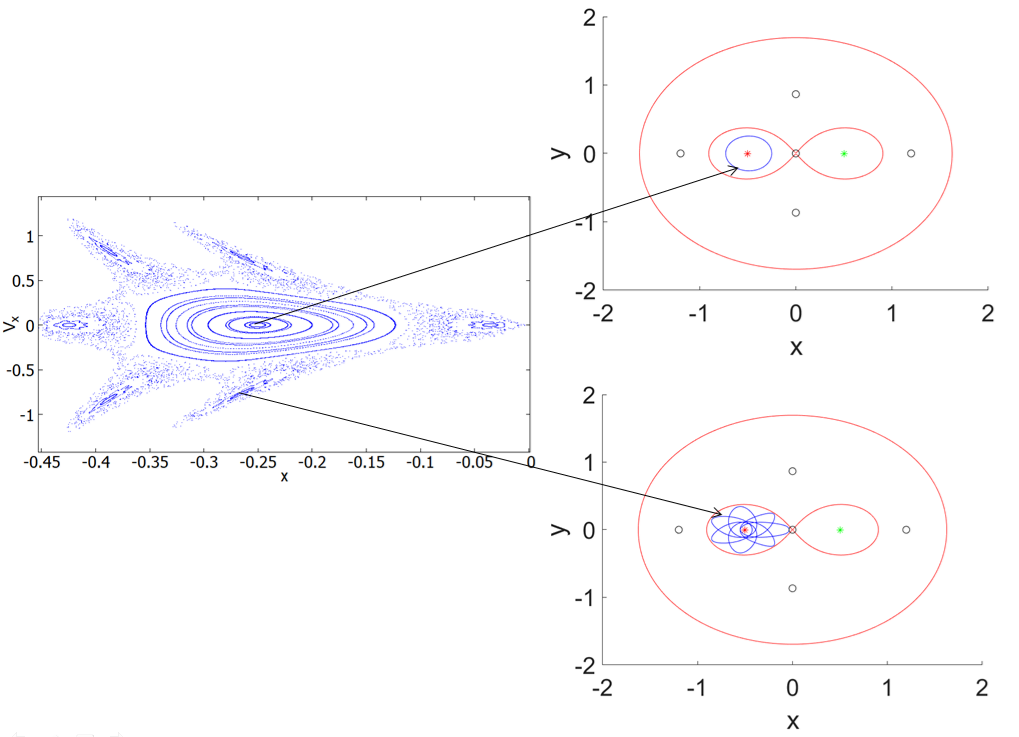}
    \caption{PSS for for $C =$ 4.00 around $M_1$ (prograde).}
    \label{1}
\end{figure}

\begin{figure}
	\includegraphics[width=\columnwidth]{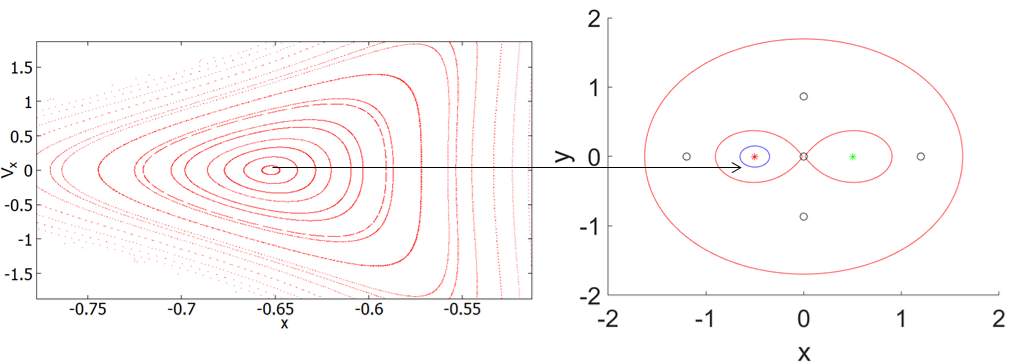}
    \caption{PSS for for $C =$ 4.00 around $M_1$ (retrograde).}
    \label{2}
\end{figure}

It is possible to observe that there are several periodic and quasi-periodic solutions around $ M_1 $ and $M_2$. Also, several points randomly distributed are present, 
which may indicate chaos. Figure \ref{1} shows a periodic solution of period 1, corresponding to a fixed central point on the $x$-axis, and also a periodic solution of period 6. The periodic orbits, around a given system, can indicate the location of possible dust around the system.
There are also periodic solutions around $ M_2 $. Figure \ref{2} shows periodic solutions of period 1 that are dominated by a fixed central point and period 6. 

These periodic orbits around $ M_1 $ and $M_2$ are surrounded by quasi-periodic orbits lying on concentric tori. In addition, we observe several isolated points in the PSS, which can be an indication of chaos. The initial conditions for some of these orbits are listed in Table \ref{tab1}. 


\begin{figure}
	\includegraphics[width=\columnwidth]{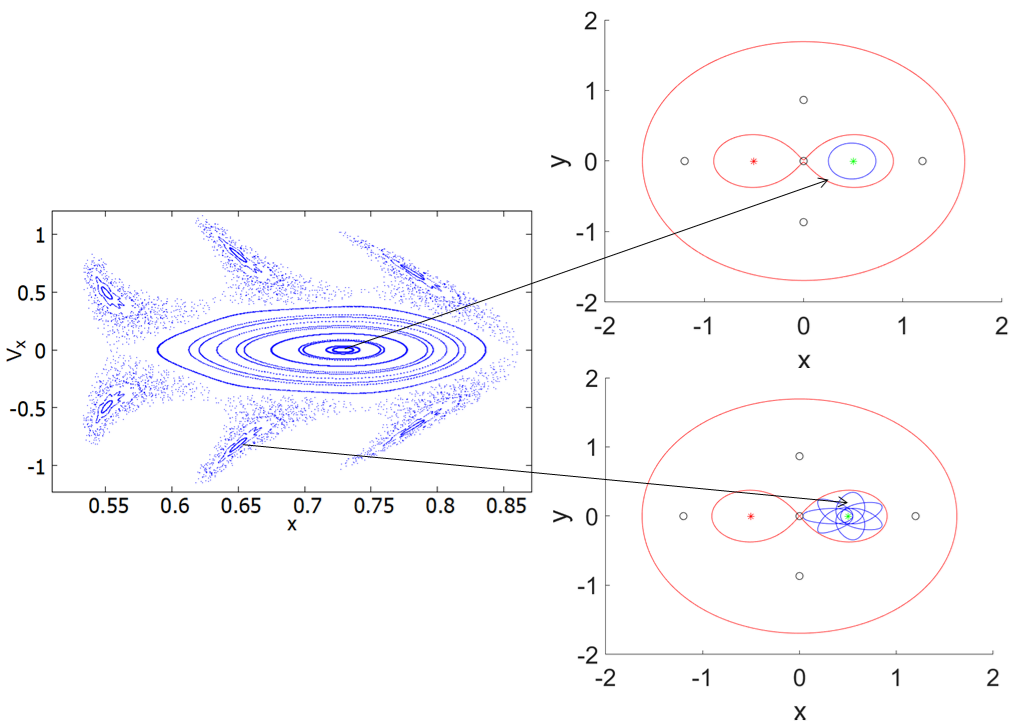}
    \caption{PSS for for $C =$ 4.00 around $M_2$ (prograde).}
    \label{3}
\end{figure}
\begin{figure}
	\includegraphics[width=\columnwidth]{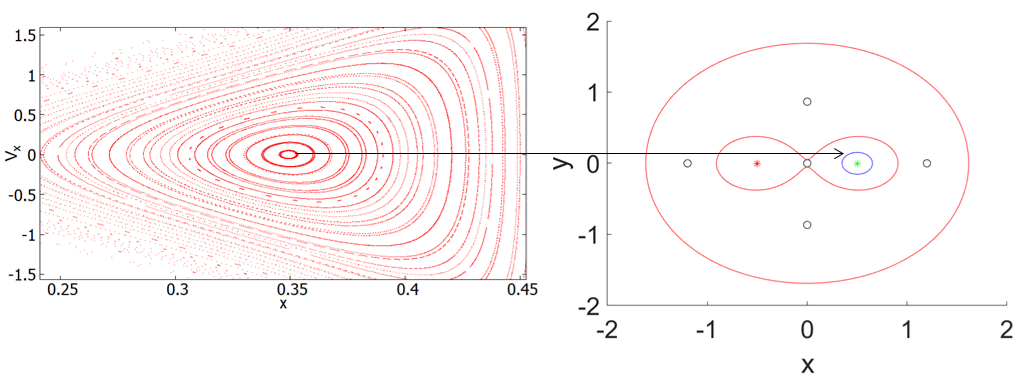}
    \caption{PSS for for $C =$ 4.00 around $M_2$ (retrograde).}
    \label{4}
\end{figure}

\begin{table}
	\centering
	\caption{Initial conditions of periodic orbits for C = 4.00.}
	\label{tab1}
	\begin{tabular}{lcccccr} 
		\hline
		$M_i-j$ & Stability & T & x & v$_y$ & $s_1$ & $s_2$ \\
		$M_1-1$ & stable & 1.326 & -0.250 & 1.185 & 1.051 & 0.303\\
		$M_1-6$ & stable & 7.844 & -0.035 & 0.145 & 1.442 & 1.225\\
	    $M_2-1$ & stable & 1.326 & 0.25 & -1.185 & 1.051 & 0.303\\
	    $M_2-6$ & stable & 7.844 &  0.035 & -0.145 & 1.442 & 1.225\\
		\hline
	\end{tabular}
\end{table}
The first column of Table \ref{tab1} gives $ M_i-j $ where $ i $ = 1, 2 (referring to primaries) and $ j $ = 1,2,3, ..., n, refers to the period of the orbit. For example, $ M_1-6 $ represents a periodic orbit around $ M_1 $ with period 6. The stability index appears in columns six and seven in Table \ref{tab1}.

\begin{figure}
	\includegraphics[width=\columnwidth]{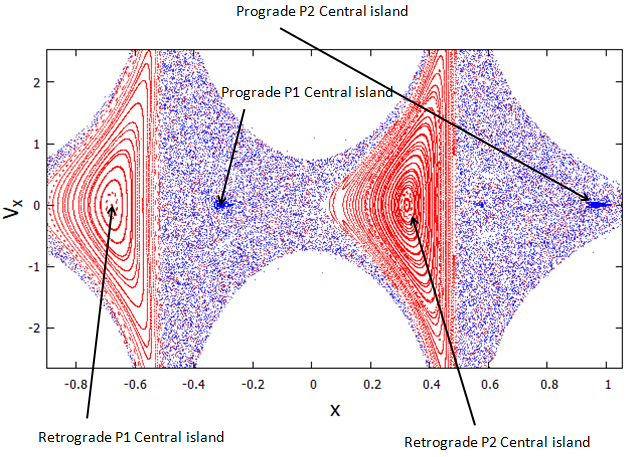}
    \caption{PSS for C = 3.4567 around 2017 YE5 system.}
    \label{generalc2}
\end{figure}

Next, we calculate the PSS considering the Jacobi constant $ C_2 $ = $ C_3 $ = 3.456796224086153.
They are shown in Figure \ref{generalc2}.

As shown in Figure \ref{generalc2}, stable retrograde islands (red) are larger compared to stable prograde islands (blue) that are inserted in a chaotic sea.

Note that, for $ C $ = $ C_2 $ = $ C_3 $, it is possible to find periodic orbits of period 1, both in direct and retrograde senses. Since the retrograde trajectories emerge on islands related to a period-1 orbit, they are usually less sensitive to perturbations with respect to the prograde trajectories.

\begin{figure}
	\includegraphics[width=\columnwidth]{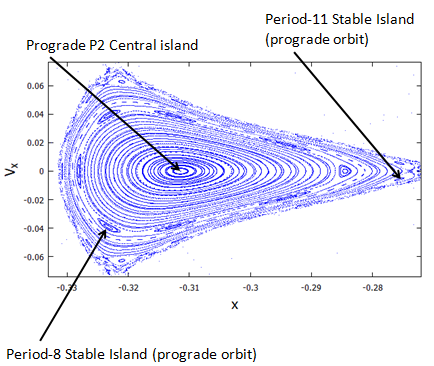}
    \caption{PSS for C = 3.4567 around $M_1$.}
    \label{5}
\end{figure}

\begin{figure}
	\includegraphics[width=\columnwidth]{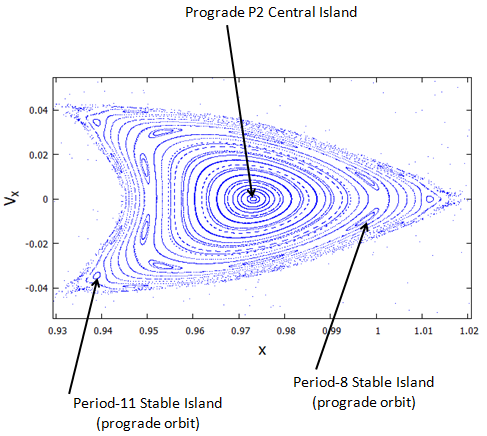}
    \caption{PSS for C = 3.4567 around $M_2$.}
    \label{6}
\end{figure}

Figures \ref{5} and \ref{6} are the detalied PSS of the prograde orbits in the vicinity of the bodies $ M_1 $ and $ M_2 $, respectively. Observe in Figure \ref{5} that the prograde motion around $M_1$ exhibits significantly-sized higher-order behavior. This means that there is a stable period-8 island and a stable period-11 island that were separated from the prograde central island. Trajectories with periods 8 and 11 are also found around $M_2$, as can be seen in Figure \ref{6}.

Note that, like in the previous case, periodic orbits are surrounded by quasi-periodic orbits on concentric circles. Here we also observe several points randomly distributed, as mentioned earlier, which may indicate chaos. 

\begin{figure}
    \includegraphics[width=1.1\linewidth]{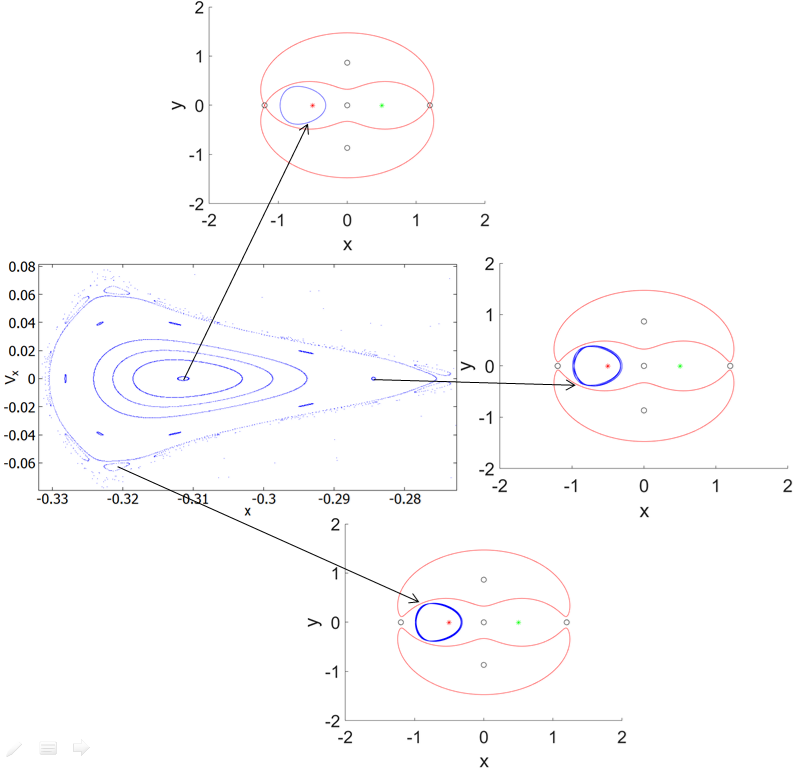}
    \caption{PSS for C = 3.4567 around $M_1$.}
    \label{direm1c2}
\end{figure}

Figure \ref{direm1c2} shows stable periodic orbits and ZVC, considering C = 3.4567, referring to the PSS of the prograde orbits around $M_1$.
 We will not plot the orbits around $M_2$ for C = 3.4567, because the orbits are similar to the orbits around $M_1$.

The initial conditions for some of these orbits are listed in Table \ref{tab2}.

\begin{table}
	\centering
	\caption{Initial conditions of periodic orbits for C = 3.45991779618}
	\label{tab2}
	\begin{tabular}{lcccccr} 
		\hline
		$M_i-j$ & Stability & T & x & v$_y$ & $s_1$ & $s_2$ \\
		$M_1-1$ & stable & 3.458 & -0.973 & -0.531 & 1.051 & 0.303\\
		$M_1-8$ & stable & 17.58 & -0.328 & 1.918 & 1.965 & 1.782\\
		$M_1-11$ & stable & 25.19 & -0.330 & 1.942 & 1.996 & 1.734\\
	    $M_2-1$ & stable & 3.65 & 0.322 & 1.872 & 1.779 & 1.753\\
	    $M_2-8$ & stable & 17.58 & 0.328 & -1.918 & 1.965 & 1.782\\
		$M_2-11$ & stable & 25.19 & 0.330 & -1.942 & 1.996 & 1.734\\
		\hline
	\end{tabular}
\end{table}

\begin{figure}
	\includegraphics[width=\columnwidth]{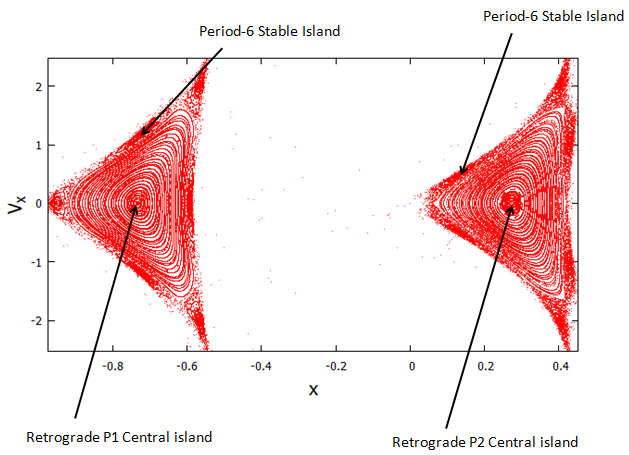}
    \caption{PSS for $C_{4-5}$ = 2.75 around 2017 YE5 system.}
    \label{generalc4}
\end{figure}

Finally, we built the PSS considering the Jacobi constant $C_{4-5}$ = 2.75, as can be seen in Figure \ref{generalc4}.

Observe in Figure \ref{generalc4} that there is no stable prograde orbit around each primary of the 2017 YE5 system when $C_{4-5}$ = 2.75.
On the other hand, retrograde orbits still exist.

\begin{figure}
	\includegraphics[width=\columnwidth]{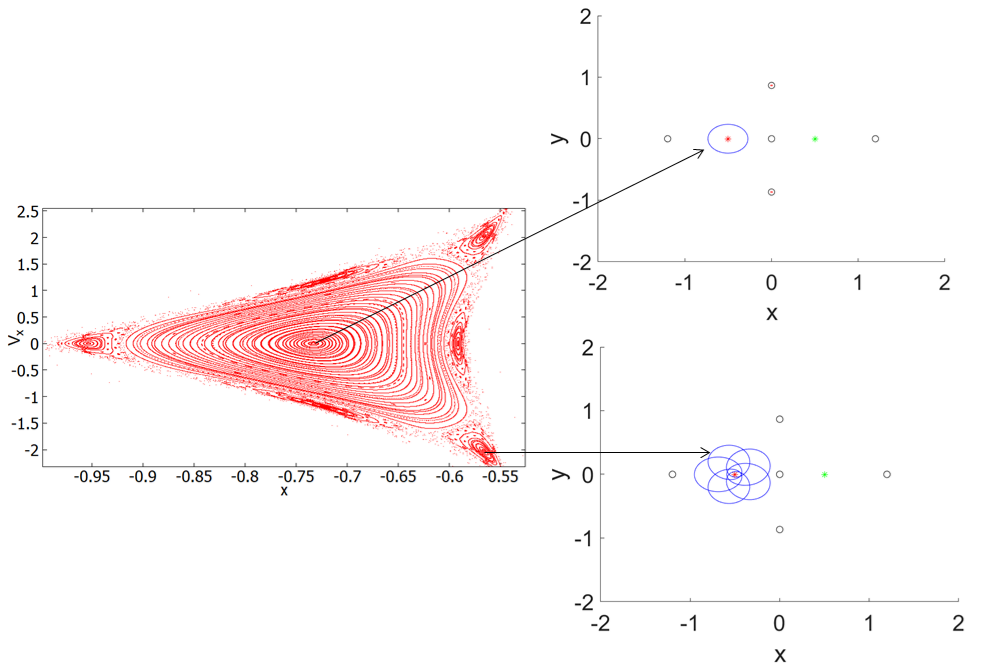}
    \caption{PSS for $C_{4-5}$ = 2.75 around $M_1$ (retrograde).}
    \label{c4m1}
\end{figure}

\begin{figure}
	\includegraphics[width=\columnwidth]{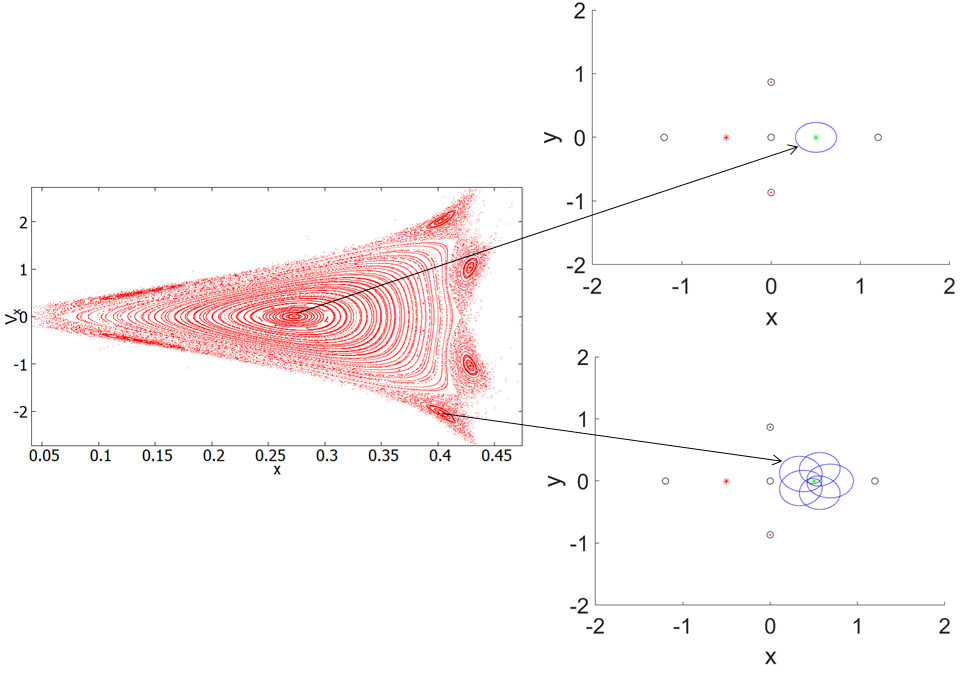}
    \caption{PSS for $C =$ 2.75 around $M_2$ (retrograde).}
    \label{c4m2}
\end{figure}

Figures \ref{c4m1} and \ref{c4m2} are the detailed PSS for retrograde orbits in the vicinity of the bodies $ M_1 $ and $ M_2 $, respectively. Note that, as in the previous cases, there is a fixed central point around of each primary. In addition, we also detect higher-order behavior around each primary. Observe that there is a stable period-6 island that was separated from the retrograde central island.

On the right side of Figures \ref{c4m1} and \ref{c4m2} we can observe the trajectories referring to the orbits of period-1 and period-6 around the bodies $ M_1 $ and $ M_2 $, respectively.

Although a large computational effort is required to generate the PSS, they are an invaluable tool for determining quasi-periodic and periodic motions.

We know that CRTBP provides the ideal periodic orbits. However, considering additional forces present in a realistic space environment, periodic orbits no longer remain periodic. In a higher-fidelity model, forces, such as SRP and 3rd body effects, usually produce quasi-periodic motion. This is due to the fact that other celestial bodies produce perturbation along the orbit that cannot be negligible. In Subsection \ref{SRP}, we will investigate the influence of solar radiation pressure on the periodic orbits found in this Subsection.

\subsection{Variation of velocity required to open transfers between different regions}
\label{VVT}

Due to the great interest of the scientific community in exploring asteroids and comets, we calculate the variations of velocity necessary for a spacecraft, close to one of the primaries, to perform a transfer between the primaries or escape from the system using (approximately) minimal energy. In this section, we seek to develop a general study of the problem, but adopting the physical characteristics of the 2017 YE5 system as a basis to develop this study. In this section, we consider the primary bodies as point masses.

\begin{figure}
\centering
    \includegraphics[width=0.7\columnwidth]{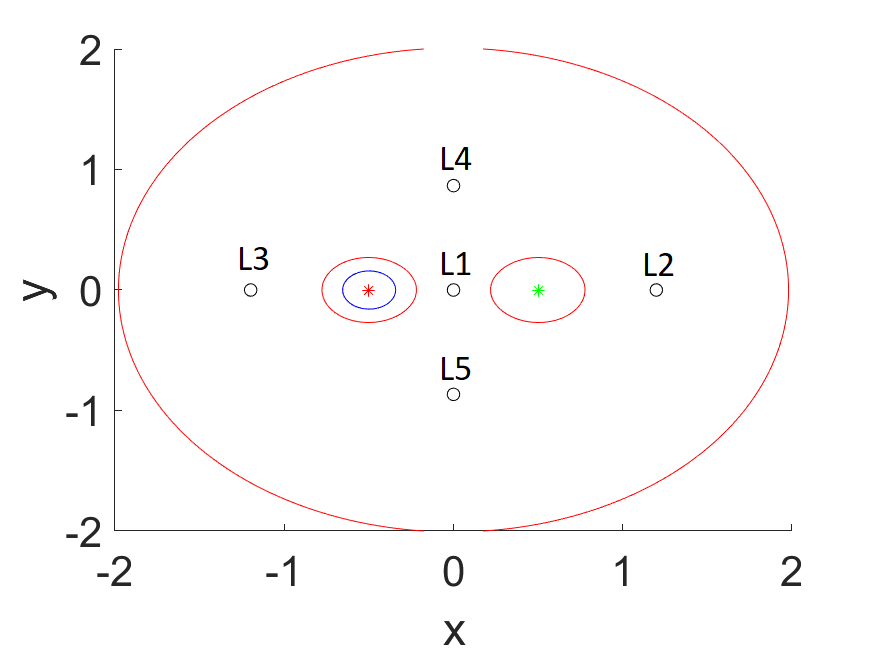}
    \caption{Periodic orbit around $M_1$ for $C$ = 5 (prograde orbit).}
    \label{c5}
\end{figure}
Figure \ref{c5} shows a periodic orbit (blue circle) around $M_1$. The Jacobi constant used is $ C = 5.00 $. Figure \ref{c5} shows that a natural orbital transfer between $ M_1 $ and $ M_2 $ is not possible, since the spacecraft would have to exceed the zero velocity curve (red circle). For this reason, it is not possible for the spacecraft to escape the system naturally.

Some orbital characteristics of the orbits shown in Figure \ref{c5} can be found in Table \ref{c55}.
\begin{table}
	\centering
	\caption{Initial conditions of periodic orbits for $C$ = 5.}
	\label{c55}
	\begin{tabular}{lcccccr} 
		\hline
		$M_{i}-j$ & Stability & T & x & v$_y$ & $s_1$ & $s_2$ \\
		$M_{1}-1$ & stable & 1.220 & -0.342 & 1.622 & 0.814 & 0.607\\
			\hline
	\end{tabular}
\end{table}

From the orbit found in Figure \ref{c5}, we want to find the $ \Delta V $ required to reach the minimum energy that allows a transfer between the primaries using an increase in tangential velocity (considering $y$ = 0 and v$_y > 0$), i. e., the $ \Delta V $ necessary for the ovals (ZVC) around $ M_1 $ and $ M_2 $ to touch each other at $ L_1 $. $\Delta V$ is the variation of the velocity applied to the initial conditions to increase the energy of the spacecraft. This velocity increment can be applied at any point in the orbit and in any direction, but the objective here is to have a quantitative assessment of the necessary fuel consumption, based on the $ \Delta V $, before carrying out a more detailed simulation. 
Consider the planar motion where the spacecraft is at position $y = 0 $ and $ x = x_0$. If we know the numerical value of the Jacobi constant, we can use Equation \ref{Omega} to find the $ v_{y_{c_i}} $ needed to open the pass-through $ L_i $ ($i$ = 1, 2, 3, 4 and 5). $ v_{y_{c_i}} $ means the velocity $ v_y $ perpendicular to the $ x$-axis whose Jacobi constant is $C_i$. Obtaining this value, we can find the $ \Delta V $ making $ \Delta V $ = $v_{y_{c_1}} $ - $v_{y_{c_5}} $.

\begin{figure}
	\includegraphics[width=\columnwidth]{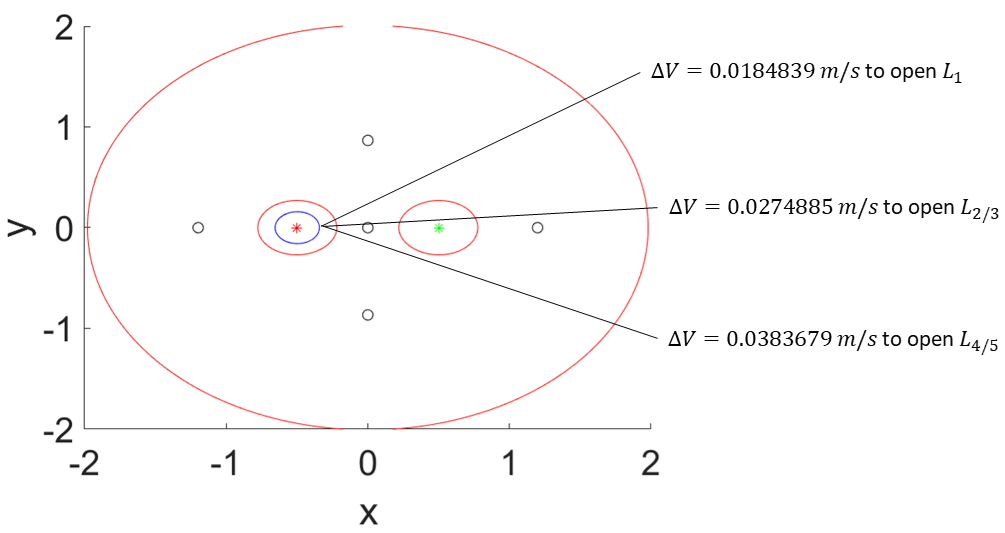}
    \caption{$ \Delta V $ required to open the passage by the Lagrangian points $ L_1 $, $ L_{2/3} $ and $ L_{4/5} $ assuming $ C = 5 $.}
    \label{deltav}
\end{figure}

In this way, we calculated the $ \Delta V $ necessary to open the equilibrium points $L_1$, $L_ {2/3}$ and $L_{4/5}$ from C = 5.0. The results are shown in Figure \ref{deltav}.

The required $ \Delta V $ depends on the position of the spacecraft in its orbit around the primaries. Note that, since this orbit is approximately circular (eccentricity = 0.0095), the $ \Delta V $ required is approximately the same at any point in the orbit, considering tangential to be the increase in velocity in the trajectory.
.

Note that the $\Delta V$ needed to open the passage by $L_{4/5}$ Lagrangian points is larger than the one to open the passage by $L_{2/3}$ points. This is due to the fact that, when the $L_{4/5}$ equilibrium points are opened, the motion of the spacecraft is allowed throughout the region, requiring a larger energy.

\begin{figure}
	\includegraphics[width=\columnwidth]{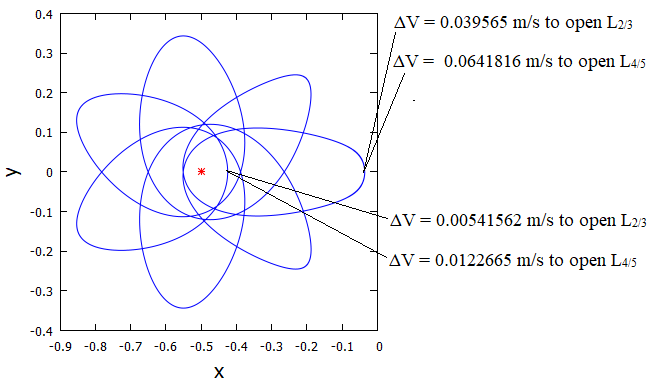}
    \caption{$ \Delta V $ required to open the passage by the equilibrium points $ L_1 $, $ L_{2/3} $ and $ L_{4/5} $ assuming $ C = 4.00 $ and using the orbit of period-6.}
    \label{deltav2}
\end{figure}

We performed the same calculations considering the period 6 advanced orbit, shown in Figure \ref{deltav2}. A closer view to $ M_1 $ and the $ \Delta V $ required to open the passage by the equilibrium $ L_{2/3} $ and $ L_{4/5} $ points are shown in Figure \ref{deltav2}.

The spacecraft is closer to $M_1$, which is a fixed point, the lower the $ \Delta V $ needed to open the subsequent equilibrium points. This result is of great importance since the amount of fuel available in a spacecraft is essential in a space mission.

\subsection{Effect of the solar radiation pressure}
\label{SRP}

After performing an analysis of the dynamics of a particle in the vicinity of the 2017 YE5 system, considering only the gravitational forces of the system, we will investigate the influence of the SRP. 

The equations of motion that we use in the numerical simulations consider a rotating system where the origin of the reference system coincides with the barycenter of the 2017 YE5 asteroid system. The equations of motion used is shown by Equation \ref{eq234}.
\begin{eqnarray} \label{eq234}
   \ddot{\textbf{r}}&=&-2\boldsymbol{\omega} \boldsymbol{\times} \dot{\textbf{r}} - \boldsymbol{\omega} \boldsymbol{\times} (\boldsymbol{\omega} \boldsymbol{\times} \textbf{r})+U_{\textbf{r}} + \mathcal{A}(\boldsymbol{P}_{rad})
\end{eqnarray}
where $\mathcal{A}$ is an instantaneous rotation that takes the vector $\boldsymbol{P}_{rad}$ from an inertial frame into a body-fixed frame. $P_{rad}$ is the acceleration due to the SRP, as presented in \citet{2002ApMRv..55B..27M, 2005mcma.book.....B}:
\begin{eqnarray}
\label{eq30}
\boldsymbol{P}_{rad} = C_r\frac{A}{m}P_s\frac{r_0^2}{r'^2_{sun}}\boldsymbol{\hat{r}},
\end{eqnarray}
where $C_r$ is the radiation pressure coefficient \citep{2017Ap&SS.362..187O, 2018Ap&SS.363...14S}. If $C_r$ = 1, it means that all radiation is absorbed by the spacecraft and, consequently, all force is transmitted to the spacecraft. On the other hand, if $C_r$ = 2, it means that all radiation is reflected and, due to the conservation of momentum, twice the force is transmitted to the spacecraft. In this work, we consider $C_r$ = 1.5. $m$ is the mass of the particle, and $A$ is the area of the cross section of the particle illuminated by the Sun. $P_s$ is the SRP at the Sun-Earth distance and its value is approximately $4.55\times10^{-6} N/m^2$; $r_0$ is the distance between the Sun and the Earth; $\hat{r}$ is the radial unit vector of the Sun with respect to the particle; $r'_{sun}$ is the Sun-particle distance. The angular velocity is 4.2120971e-08 rad/s, and its orbital period around the Sun is $T$ = 4.73 years. We performed numerical simulations considering two situations: the asteroid at perihelion (0.818 au) and the asteroid at aphelion (4.818 au) of the orbit. In this work, we used three scenarios for the SRP: no SRP; area-to-mass ratio equal to 0.01 and 0.1. In this form, we can infer the effect for a large range of spacecraft. 

When the 2017 YE5 system is in the perihelion, we observe that the SRP significantly perturbs the original CR3BP periodic solutions when it is included. None of the orbit geometries are maintained when SRP is included.

\begin{figure}
	\includegraphics[width=\columnwidth]{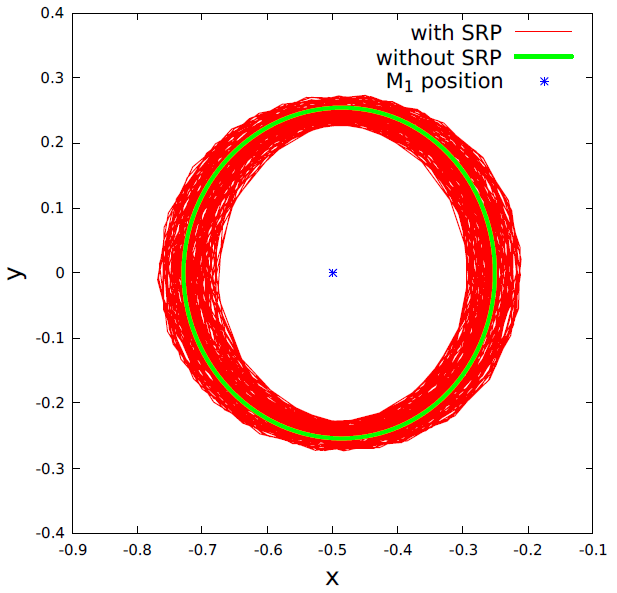}
    \caption{Orbit period-1 around $M_1$ considering SRP using $C = 4.00$.}
    \label{SRP1}
\end{figure}

On the other hand, when we consider $A/m = 0.01$, and the asteroid system 2017 YE5 at aphelion, it is possible to find orbits that survive around each primary body. Figure \ref{SRP1} shows an orbit around $M_1$ considering the SRP (red) and when we do not consider the SRP (green). Note that when SRP is taken into account, the orbit oscillates around the periodic orbit period-1. In this simulation, we consider the Jacobi constant $C = 4.00$.

\begin{figure}
	\includegraphics[width=\columnwidth]{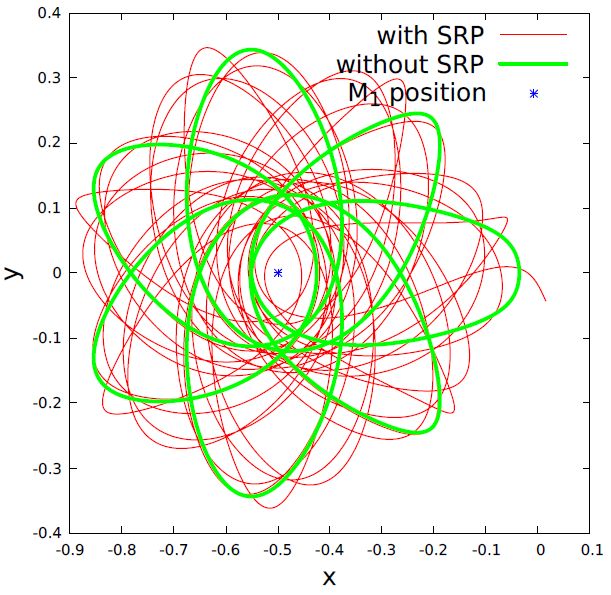}
    \caption{Orbit period-6 around $M_1$ considering SRP using $C = 4.00$.}
    \label{SRP2}
\end{figure}

In the numerical simulations that we performed (for 1 year), we observed that the effect of the SRP, when the 2017 YE5 system is at aphelion and $A/m = 0.01$, is not enough to make the period 1 orbit escape from the system or orbit the other primary.
On the other hand, the SRP has a larger influence on the orbits with a period longer than one, causing the orbit to escape from the vicinity of one of the primaries, as can be seen in Figure \ref{SRP2}.

Note in Figure \ref{SRP2} that, when we consider the SRP, the orbit (red) oscillates  around the periodic orbit (green). After a few days, the red orbit escapes from the vicinity of $M_1$ (through the equilibrium point $L_1$, coordinate (0,0) in the figure) and approaches $M_2$. As we consider the Jacobi constant $C = 4.00$, this path is the only one that allows orbital transfers between the primaries. In Figure \ref{SRP2}, we show the trajectory until the moment when the orbit escapes through the $L_1$ point, to avoid visual pollution.

Because the orbits with a period $>$ 1 have a higher stability index with respect to the orbits of period 1, they are more susceptible to disturbances from other celestial bodies.

\begin{figure}
	\includegraphics[width=\columnwidth]{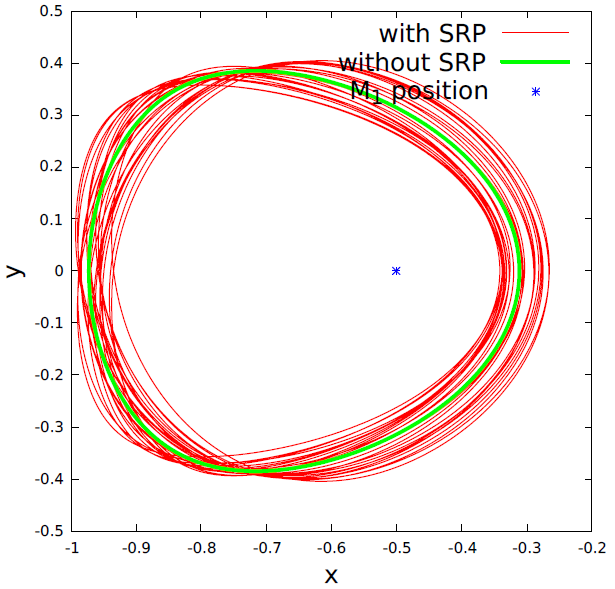}
    \caption{Orbit period-1 around $M_1$ considering SRP using $C = 3.4567$.}
    \label{SRP33}
\end{figure}

In Figure \ref{SRP33}, we consider C = 3.4567. This figure shows an orbit around $M_1$ considering the SRP (red) and when we do not consider the SRP (green). Note that, when SRP is taken into account, the orbit oscillates around the periodic orbit period-1. This simulation was carried out during 1 year of numerical integration, assuming that the 2017 YE5 system is in its aphelion.

Similar results were found when considering other Jacobi constants and orbital periods.

\begin{figure}
\centering
    \includegraphics[width=0.9\columnwidth]{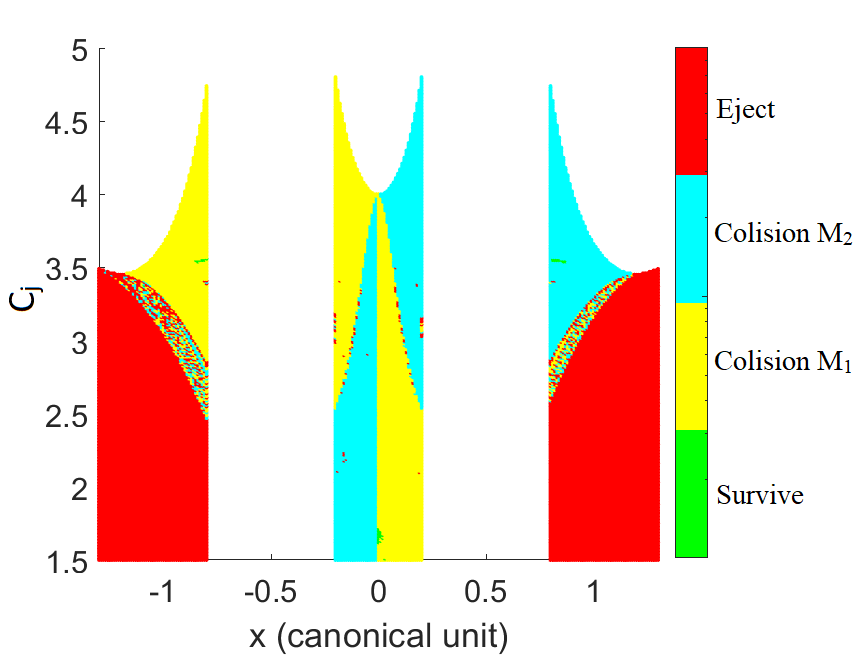}
    \caption{Initial conditions grid of the direct orbits around the 2017 YE5 system. Taking into account the SRP and the size of the primary bodies.}
    \label{orbitmapdirect}
\end{figure}

\begin{figure}
\centering
	\includegraphics[width=0.9\columnwidth]{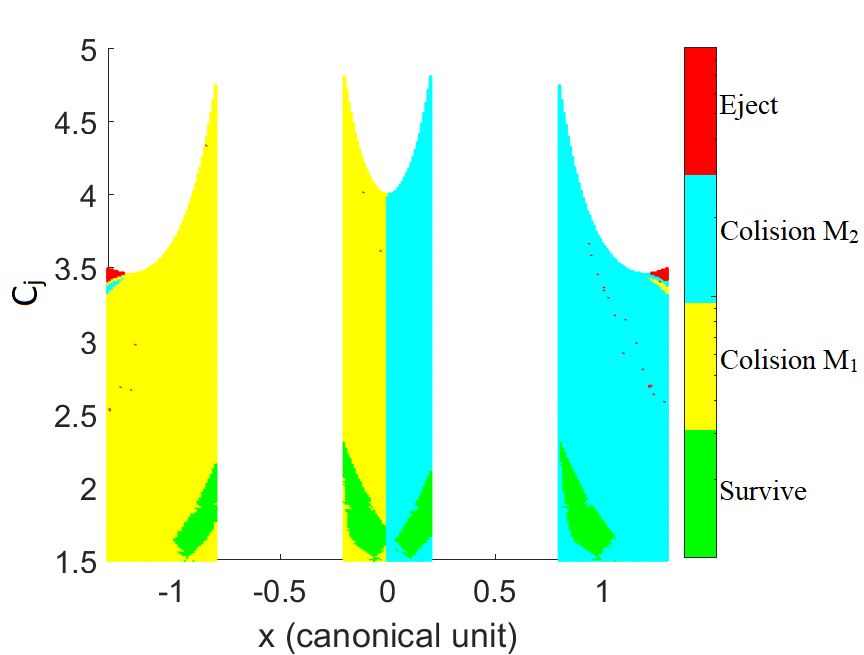}
    \caption{Initial conditions grid of the retrograde orbits around the 2017 YE5 system, taking into account the SRP and the size of the primary bodies.}
    \label{orbitmapretrograde}
\end{figure}

In Figures \ref{orbitmapdirect} and \ref{orbitmapretrograde}, we present the initial conditions that generate direct and retrograde orbits, respectively, colliding with $M_1$ (yellow points), colliding with $M_2$ (cyan points), orbiting the system (green points), and escaping the system (red points). We assume that the orbit escapes the 2017 YE5 system when the particle is at the position $R_{hill}$ + 50 km away from the primary bodies, where $R_{hill}$ is the sphere of influence of the 2017 YE5 system (100 km). Collision with the primary bodies occurs when the particle crosses the radius of the primary bodies ($r_1$ = $r_2$ = 450 m). The numerical simulations were performed when the particle (massless body) is initially at a position $>$ 0.055 canonical unit (100 meters) away from the surface of the primary bodies. Thus, the white areas are regions very close to the surface of the primary bodies or forbidden regions, limited by the zero velocity curves. In the direct orbit diagram, smooth and fractal-like boundaries between the sets may be observed \citep{Assis_Terra}; while boundaries are mostly smooth for the retrograde orbit diagram.

Due to the fact that the primary bodies have equal masses and are spherical in shape, it is possible to observe some symmetries in the initial condition grids. If not for the Sun's perturbation, the initial conditions grid of the $x$ axis $>$ 0 would be a mirror of the grid when $x$ $<$ 0.

Numerical evidence shows that no orbit ejects (both in direct or retrograde orbits) from the system when the Jacobi constant $C_j$ $>$ 3.5. We observe that, when we do not consider the perturbations, the equilibrium points $L_2$ and $L_3$ open to $C_j$ = 3.4567. Although the SRP increases the energy of the system over time (decreases the forbidden regions), values of $C_j$ $<$ 3.5 are necessary for the escapes to occur.

It is important to note that we found orbits that survived around the system when $C_j$ = 4.0 (Figs. \ref{SRP2} and \ref{SRP33}). These orbits survive when we do not take into account the size of the primary bodies. On the other hand, when we consider the dimension of the primary bodies, Figs. \ref{orbitmapdirect} and \ref{orbitmapretrograde} shows that no orbit survives for this energy. The orbits survive through all numerical integration from $C_j$ = 3.55 only for direct orbits. 
For $C_j$ $<$ 3.5, the direct orbits practically escape from the system (red regions) or collide with one of the primary bodies (yellow or cyan regions).
On the other hand, for $C_j$ $<$ 3.5, it is possible to find several initial conditions that make the orbits survive through all numerical integrations.
Naturally, space dust particles exist in the vicinity of some celestial bodies in direct orbits. In contrast, retrograde orbits hardly arise naturally. Thus, regions where direct orbits do not survive and retrogrades do survive are great options for inserting a spacecraft, because in these regions, the probability of existing space dust particles is low, reducing the risk of a spacecraft colliding with some space object.

\begin{figure}
	\includegraphics[width=\columnwidth]{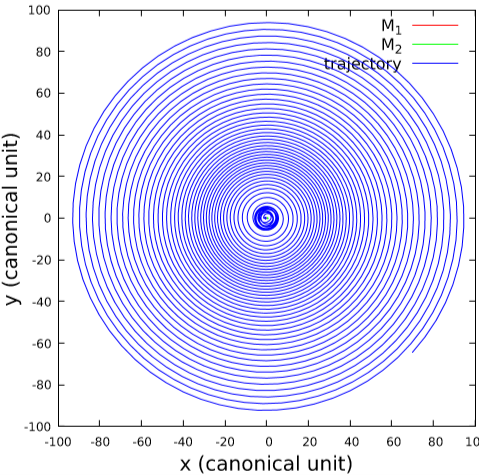}
    \caption{System ejecting orbit (blue). The initial position of the Sun is on the right side of the Figure. The dimensions of $M_1$ and $M_2$ are shown in red and green, respectively. The SRP was taken into account. Here we consider $C_j$ = 3.41 and $x$ = -0.1799.}
    \label{C0.341X}
\end{figure}

\begin{figure}
	\includegraphics[width=\columnwidth]{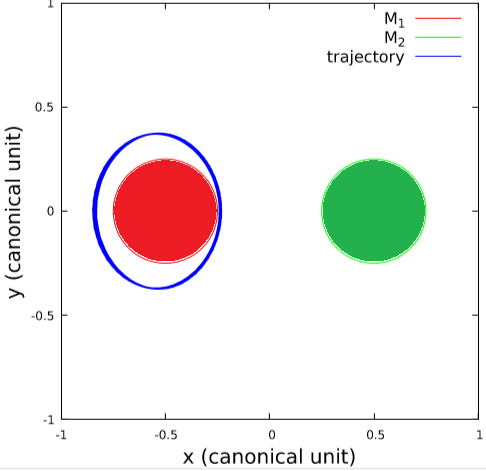}
    \caption{Particle orbiting $M_1$ (blue). The initial position of the Sun is on the right side of the Figure. The dimensions of $M_1$ and $M_2$ are shown in red and green, respectively. The SRP was taken into account. Here we consider $C_j$ = 3.55 and $x$ = -0.2299.}
    \label{C355X}
\end{figure}

\begin{figure}
\includegraphics[width=\columnwidth]{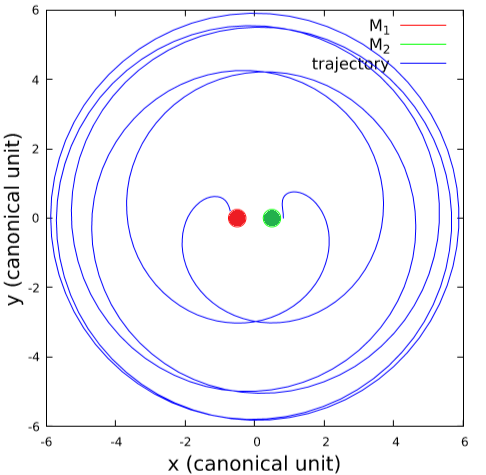}
    \caption{Orbit (blue) starts near $M_2$, circles the system and collides with $M_1$. The initial position of the Sun is on the right side of the Figure. The dimensions of $M_1$ and $M_2$ are shown in red and green, respectively. The SRP was taken into account. Here we consider $C_j$ = 2.73 and $x$ = 0.8301.}
    \label{C273X0}
\end{figure}

\begin{figure}
\includegraphics[width=\columnwidth]{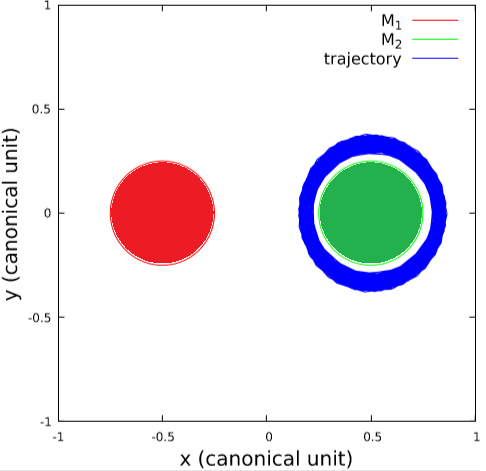}
    \caption{Particle orbiting $M_2$ (blue). The initial position of the Sun is on the right side of the Figure. The dimensions of $M_1$ and $M_2$ are shown in red and green, respectively. The SRP was taken into account. Here we consider $C_j$ = 2.00 and $x$ = 0.2001.}
    \label{C2X0}
\end{figure}

Extra examples of our results for direct orbits can be found in Figures \ref{C0.341X} and \ref{C355X}, with ($C_j$, $x$) = (3.41, -0.1799) and (3.55, -0.2299) whereas, for retrograde orbits, the extra examples can be found in Figures \ref{C273X0} (2.73, 0.8301) and \ref{C2X0} (2.0, 0.2001), respectively.

\section{Transfers between collinear points}
\label{transfer}

There are several publications where orbital maneuvers between the Lagrange points are studied. Among these publications, we can cite the problem of orbital maneuvers between the Lagrange points of the Sun-Earth system \citep{1979JGCD....2..257B}, the Earth-Moon system \citep{1996AcAau..39..483P} as well as the Sun-Earth-Moon system \citep{2008AcAau..63.1221C}. \cite{2015Ap&SS.357...66Y} investigated the problem of orbit transfers connecting irregularly shaped asteroid equilibrium points. In the study by \cite{2015Ap&SS.357...66Y}, only the gravity of the asteroid in its dynamics was considered. The gravitational influence coming from other celestial bodies was considered to be very small in relation to the gravitational forces of point-mass asteroids.
The work done by \cite{2017Ap&SS.362..187O} investigated the orbital transfers between the equilibrium points of a binary asteroid system, taking into account the solar radiation pressure.

In this section, we will investigate orbital transfers between the equilibrium points of the 2017 YE5 system. In addition to the gravitational effect of the primary bodies, we take into account the SRP. We want to investigate the $\Delta$V required for a spacecraft to move from one equilibrium point to another. We assume that, when a spacecraft is at an equilibrium point, the solar panel is parallel to the sunlight, in such a way that the influence of SRP does not change the position of the equilibrium points, i.e., SRP only disturbs the spacecraft during the transfer trajectory. The type of transfer that we will use to perform this study is the bi-impulsive maneuver. To carry out the simulations, we consider that initially ($t$ = 0) the three bodies ($M_1$, $M_2$ and the Sun) are aligned, as shown in Figure \ref{transor}.

\subsection{Numerical method}

Considering that the spacecraft is positioned at an equilibrium point in the 2017 EY5 system, it may be necessary for the spacecraft to change its orbit during a mission, such that it can study and collect information about each of the bodies in this system, for example.

\begin{figure}
	\includegraphics[width=\columnwidth]{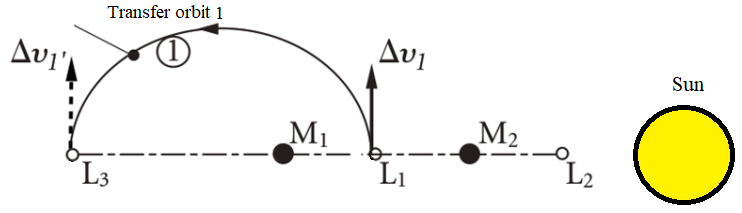}
    \caption{bi-impulsive maneuver from the equilibrium point $L_1$ to $L_3$. Source and adapted from \citet{2017Ap&SS.362..187O}.}
    \label{transor}
\end{figure}

Figure \ref{transor} shows how these transfers are planned. It begins with the application of the first impulse at the initial position of the spacecraft, which is initially at a Lagrange point ($L_1$), and ends with the application of the second impulse in the final position of the space vehicle, the Lagrange endpoint ($L_3$). In this way, an impulsive maneuver is applied at point $L_1$, giving the space vehicle the required $\Delta$$v$$_1$ to initiate the transfer from orbit 1 towards point $L_3$. When the space vehicle reaches $L_3$, another impulsive maneuver is applied, giving the space vehicle the necessary variation of velocity $\Delta$$v$$_1'$, aiming to place the spacecraft in the desired final orbit of the Lagrange point $L_3$. Because these maneuvers require two impulses to complete the transfer, this type of maneuver is called a bi-impulsive maneuver.

The equation of motion for the CRTBP has no analytical solutions, and numerical integrations are required to be applied to solve the problem. Therefore, this problem is treated as a Two Point Boundary Value Problem (TPBVP), which is a problem in which ordinary differential equations are needed to satisfy the boundary conditions on more than one value of the independent variable.

Thus,

(i) it is necessary to provide an initial value for the position ($\vec{r}_i$) and the velocity ($\vec{v}_i$);

(ii) the final state is given by the desired final velocity $\vec{v}_d$ and the desired final position $\vec{r}_d$;

(iii) next, it is necessary to determine a final time, $\tau_f$;

(iv) then, calculate the new state, composed by a velocity vector $\vec{v}_f$ and a position vector $\vec{r}_f$, both obtained from the numerical integration of the initial state from $0$ to $\tau_f$;

(v) finally, check the final position. If $| \vec{r}_f$ - $\vec{r}_d|$ is less than a given tolerance (10$^{-5}$), the solution is found and the process is stopped. Otherwise, the process returns to step (i), and an increment in the initial velocity $\vec{v}_i$ is done.

The solution provides the spacecraft trajectory as well as fuel consumption amounts, specified by the amount of $\Delta$V over the entire transfer time, which is the $\Delta$$v$$_1$ at spacecraft launch plus the $\Delta$$v$$_1'$ at the desired endpoint. Thus, changing the flight time makes it possible to find a family of transfer orbits.

\begin{figure}
\centering\includegraphics[scale=0.5]{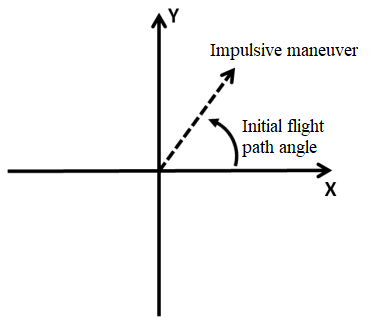}
\caption{Initial flight path angle. Source and adapted from \citet{2017Ap&SS.362..187O}.}
\label{AD}
\end{figure}

The results also show graphs with the variation of velocity $\Delta$V versus time, and the variation of velocity $\Delta$V versus the initial flight path angle (FPA), as done by \cite{1996AcAau..39..483P}. The definition of this angle is such that zero is on the $X$ axis, pointing in the positive direction and increasing counterclockwise. This definition is shown in Figure \ref{AD}.

In this work, we consider the time of flight from $\tau_f$ = 0.1 days to 1 day, i.e., up to one orbital period of the system.

\subsection{Orbital transfers}

\begin{figure}
\centering\includegraphics[scale=0.5]{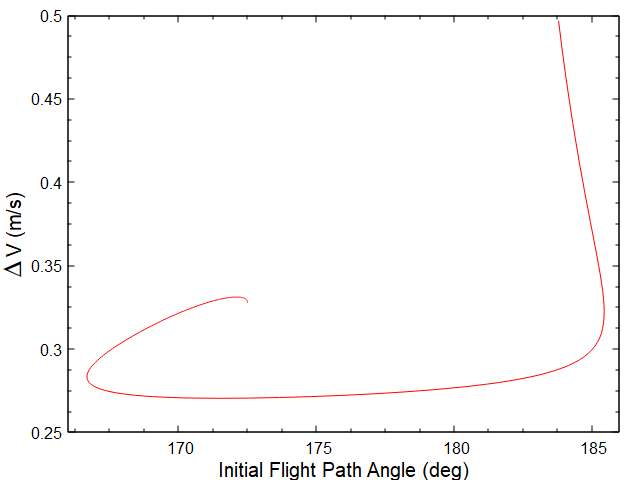}
\caption{Transfers from $L_2$ to $L_1$, giving the firing angle vs $\Delta$V.}
\label{L2toL1fp}
\end{figure}

\begin{figure}
\centering\includegraphics[scale=0.5]{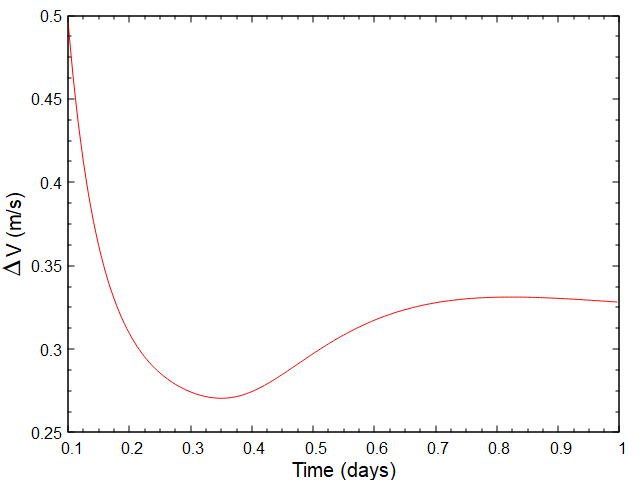}
\caption{Transfer from $L_2$ to $L_1$, giving transfer time vs $\Delta$V.}
\label{L2toL1time}
\end{figure}

Simulations of orbital transfers were carried out between the equilibrium points $L_1$ to $L_2$, $L_2$ to $L_3$, and, finally, from $L_3$ to $L_2$ of the 2017 YE5 system. As a result, the plots with the corresponding $\Delta$V as a function of the initial flight path angle, and the corresponding $\Delta$V as a function of the time are shown in Figure \ref{L2toL1fp} and Figure \ref{L2toL1time}, respectively. In all transfer simulations, we take into account the gravitational force of the primary bodies (M1 and M2), as well as the SRP.

The first transfer orbit family considers transfers between $L_2$ and $L_1$, considering that the asteroid is in the perihelion of its orbit around the Sun, which is about 0.818 au.

\begin{figure}
\centering\includegraphics[scale=0.5]{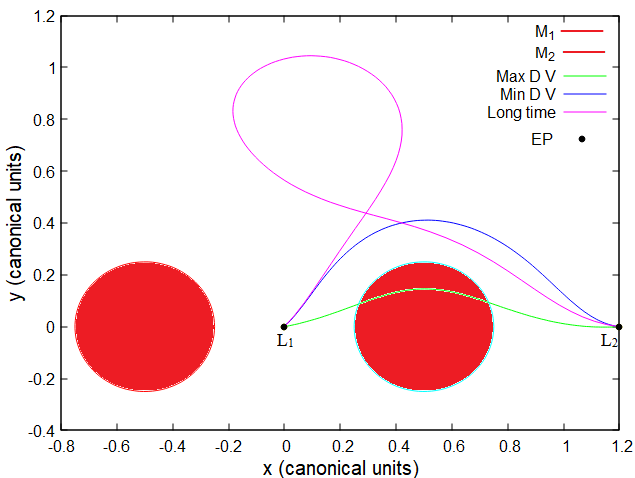}
\caption{Trajectories of a space vehicle from $L_2$ to $L_1$.}
\label{traj_l2tol1}
\end{figure}

It is possible to observe, from Figure \ref{L2toL1fp}, that the FPA responsible for the orbital transfer family from $L_2$ to $L_1$ is around 165$^\circ$ and 185$^\circ$. This is because we need to apply the impulse towards the equilibrium point $L_1$, which is 180$^\circ$ from $L_2$. Figure \ref{L2toL1time} provides information of $\Delta$V as a function of the time of flight (TOF) during the transfer. Note that, for a fast transfer (t = 0.1 days), a high $\Delta$V is required. This high value comes from the fact that the spacecraft has to travel at high speed to reach the equilibrium point in less time. As the transfer time increases, the $\Delta$V required to carry out the maneuver becomes smaller, until it reaches a minimum value ($\Delta$V = 0.276408 m/s, t = 0.35 days). This is because, as we take longer to reach the final destination, the spacecraft can travel at a slower speed, thus needing a smaller $\Delta$V. After reaching a minimum, transfer time becomes longer and $\Delta$V also increases. This is easy to understand if we look at Figure \ref{traj_l2tol1}, which shows the trajectory of the spacecraft. This figure shows the position of the primary bodies that are represented by the red circles. The trajectory shows maximum $\Delta$V is the green path; the minimum $\Delta$V in the blue path and the trajectory that tooks the longest time to get out of $L_2$ and reach $L_1$ (purple path); as well as the position of the equilibrium points (EP), shown as black points.

Note that the trajectory that takes longer time has a FPA similar to the trajectory that spends a minimum $\Delta$V (blue and purple trajectory), but the purple trajectory reaches a much higher point (with respect to the $y$ axis) when compared to the other trajectories. This is because the initial thrust is higher, making the spacecraft travel a greater distance in space and, consequently, spending more time and fuel to reach the final destination ($L_1$).

It is important to note that the trajectory that takes the shortest time (green) has FPA = 183.79$^\circ$. This trajectory seeks the shortest path to reach the equilibrium point $L_1$, consequently, the spacecraft collides with $M_2$ during this transfer.

\begin{figure}
\centering\includegraphics[scale=0.5]{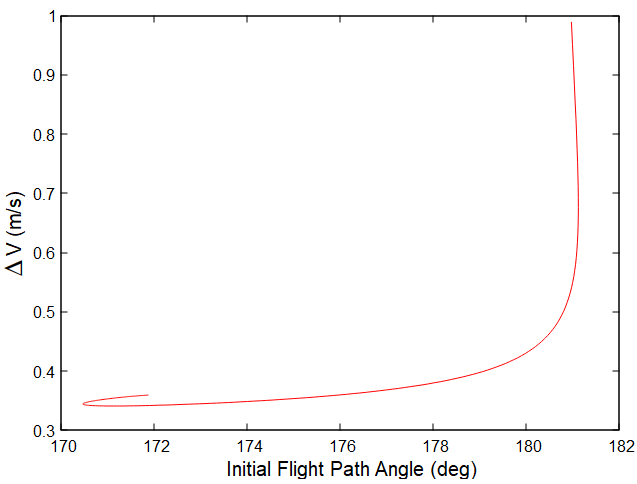}
\caption{Transfers from $L_2$ to $L_3$, giving the firing angles vs $\Delta$V.}
\label{L2toL3fp}
\end{figure}

\begin{figure}
\centering\includegraphics[scale=0.5]{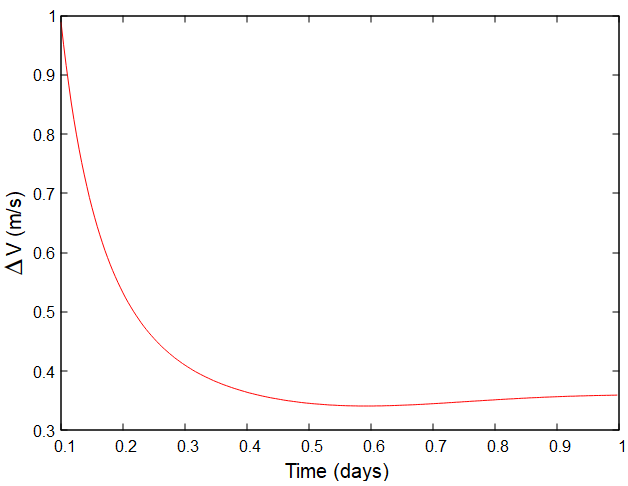}
\caption{Transfers from $L_2$ to $L_3$, giving transfer times vs $\Delta$V.}
\label{L2toL3time}
\end{figure}

Next, we performed the transfer simulations from $L_2$ to $L_3$. Figures \ref{L2toL3fp} and \ref{L2toL3time} provide information about the $\Delta$V vs initial flight path angle and the $\Delta$V vs flight time, respectively.

Note in Figure \ref{L2toL3fp} that the FPA is between 165$^\circ$ and 185$^\circ$ for the same reasons mentioned in the interpretation of Figure \ref{L2toL1fp}. Note that the shortest flight time (t = 0.1 days) uses the maximum $\Delta$V, because to reach the final destination it is necessary to have a higher speed and, therefore a higher $\Delta$V. It is also possible to notice that the $\Delta$V needed to get out of $L_2$ and reach $L_3$ in the shortest time (t = 0.1 days) is double the value found to get out of $L_2$ and reach $L_1 $ at the same time. This is due to the symmetry of the problem. Due to the fact that the primary bodies have equal masses, the equilibrium point $L_1$ is halfway from $L_3$ with respect to a space vehicle starting from $L_2$. For the spacecraft to travel twice the distance (reach $L_3$) in the same time interval as it reached $L_1$ (t = 0.1), it is necessary to double the speed value.

In Figures \ref{L2toL3fp} and \ref{L2toL3time}, there is a minimum $\Delta$V needed for the spacecraft to reach $L_3$ departing from $L_2$. The minimum $\Delta$V has the value $\Delta$V = 0.341153 m/s, the flight time is t = 0.5917 days and FPA = 171.134$^\circ$. Note that the minimum $\Delta$V to reach $L_3$ starting from $L_2$ is slightly higher than the minimum $\Delta$V to reach $L_1$ starting from $L_2$. This occurs because the equilibrium point $L_3$ is far away than $L_1$ from $L_2$ (when departure from $L_2$), requiring more energy to reach the region that is $L_3$. The flight time is also slightly longer than the transfer time to reach $L_1$ from $L_2$. This occurs because the equilibrium point $L_3$ is the farthest point from $L_2$, requiring longer to reach this point using an optimal value of $\Delta$V.

\begin{figure}
\centering\includegraphics[scale=0.5]{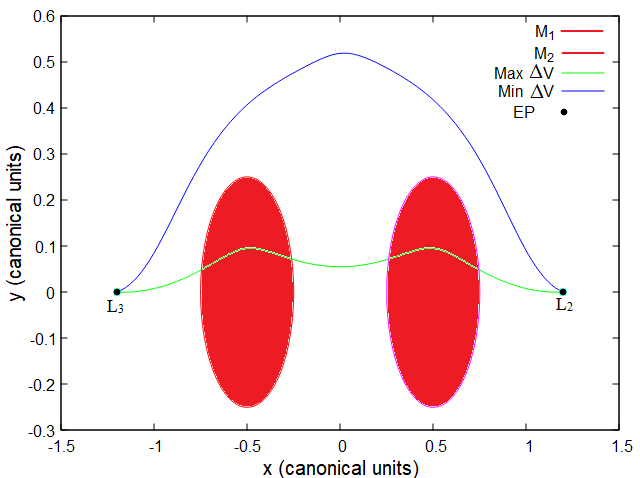}
\caption{Trajectories of a space vehicle starting from $L_2$ and reaching $L_3$.}
\label{traj_l2tol3}
\end{figure}

In Figure \ref{traj_l2tol3}, it is possible to observe the trajectory of the space vehicle starting from $L_2$ and reaching $L_3$. We plot only the optimized trajectory (minimum $\Delta$V) and the trajectory with the maximum $\Delta$V (0.1 transfer time). Note that, as in the previous case, the trajectory with the highest $\Delta$V collides with $M_2$. On the other hand, the optimized trajectory manages to reach the final destination using minimal fuel and spending only 12 hours to carry out the transfer.

\begin{figure}
\centering\includegraphics[scale=0.5]{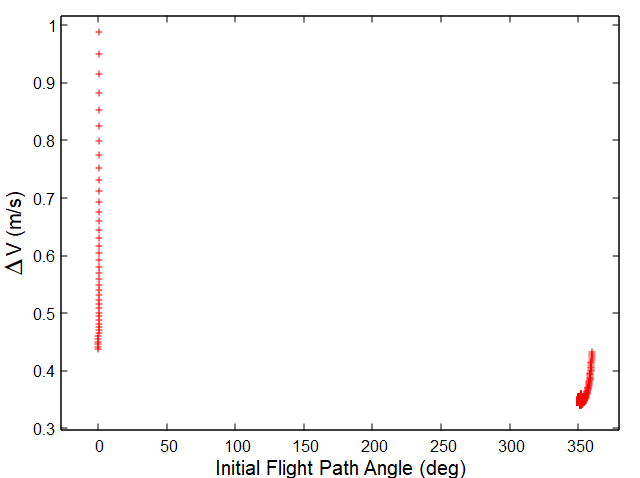}
\caption{Transfers from $L_2$ to $L_3$. FPA vs $\Delta$V.}
\label{L3toL2fp}
\end{figure}

\begin{figure}
\centering\includegraphics[scale=0.5]{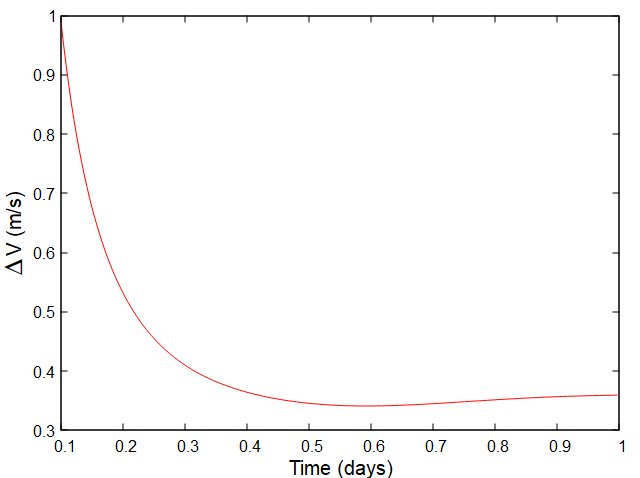}
\caption{Transfers from $L_2$ to $L_3$. Flight time vs $\Delta$V.}
\label{L3toL2time}
\end{figure}

Finally, we carry out the numerical simulations considering a space vehicle that starts from $L_3$ and reaches $L_2$. The transfer families are shown in Figures \ref{L3toL2fp} and \ref{L3toL2time}.

In Figure \ref{L3toL2fp} we can see that we find a family of orbital transfers for a FPA around 0$^\circ$, since $L_2$ makes this angle with respect to an object that is in $L_3$ (positive direction of axis $x$). Next, it is only possible to find transfer orbit families when the FPA is around 345$^\circ$ and 360$^\circ$. This discontinuity means that for FPA between 0$^\circ$ and 345$^\circ$ is not possible for a spacecraft to reach $L_2$ departing from $L_3$ for the 2017 YE5 system. The minimum amount to carry out this transfer is $\Delta$V = 0.341152 m/s with FPA = 171.13$^\circ$.

\begin{figure}
\centering\includegraphics[scale=0.5]{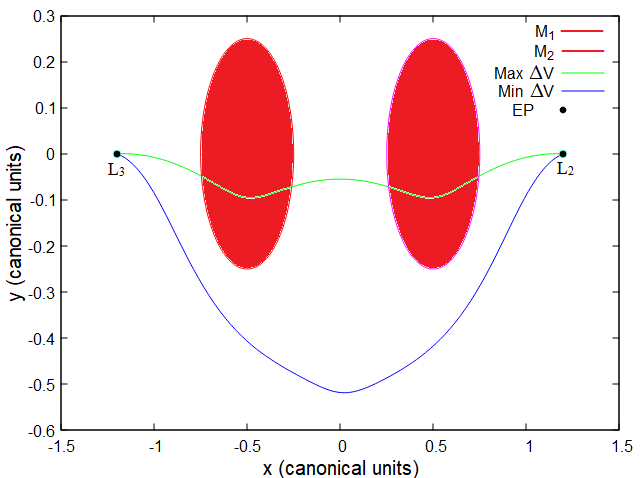}
\caption{Trajectories of a space vehicle starting from $L_3$ and reaching $L_2$.}
\label{traj_l3tol2}
\end{figure}

Figure \ref{L3toL2time} provides flight time information. Note that this Figure is very similar to Figure \ref{L2toL3time}. This is due to the symmetry of the system. As the primary bodies of the 2017 YE5 system have equal masses, the same impulse is needed to depart from the symmetric equilibrium points, which in this case is $L_3$ and $L_2$. This can be verified in Figure\ref{traj_l3tol2}.

In Figure \ref{traj_l3tol2} we plot only the optimal trajectory ($\Delta$V minimum) and the trajectory with the shortest time (t = 0.1). Note the similarity between Figures \ref{traj_l2tol3} and \ref{traj_l3tol2}. We can observe that the energy required to reach $L_2$ starting from $L_3$ is the same as the one to go the other way around (from $L_2$ to $L_3$). Because the area/mass ratio is small ($A/m$ = 0.01), the disturbance due to the SRP does not significantly influence the trajectory during the transfer, since this transfer is performed in a small time interval (maximum 1 day). The perturbation of SRP is, on average, 1000 times smaller than the gravitational forces of the primary bodies so, consequently, the gravitational forces of the primary bodies predominantly govern spacecraft motion in this region and in the flight time interval.

The difference $\Delta$V$_{L_2~to~L_3}$ - $\Delta$V$_{L_3~to~L_2}$ = 5.1 $\times$ $10^{-8}$ m/s, where $\Delta$V$_{L_i~to~L_k}$ means the $\Delta$V needed to get out of $L_i$ and reach $L_k$. Note that this difference is low but still non-zero. If we do not consider the disturbance from the Sun, this difference would be zero.

\begin{figure}
\centering\includegraphics[scale=0.5]{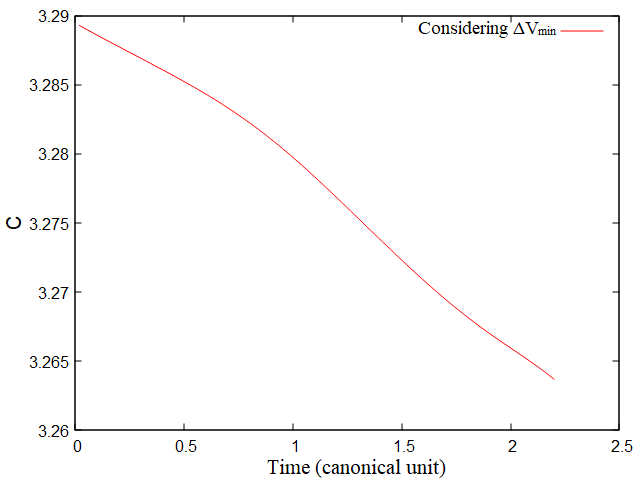}
\caption{Jacobi constant versus transfer time during the transfer from $L_2$ to $L_1$ with $\Delta$V minimum.}
\label{cl2l1}
\end{figure}

Figure \ref{cl2l1} provides information on the Jacobi constant C vs time considering the optimal $\Delta$V of the transfers from $L_2$ to $L_1$. The figure only shows the value of the Jacobi constant during the transfer path, i.e., after the initial impulse and before the final impulse. Note that the ``constant'' of Jacobi C changes as time passes. This is due to the fact that the Sun's perturbation causes the space vehicle to gain energy, thus decreasing the value of the Jacobi constant at each instant of time. As the transfer time is short, the Jacobi constant does not vary considerably during the transfer ($\Delta$C $\approx$ 0.026).
If we do not consider the disturbance from the Sun, C would be constant throughout the entire transference trajectory.

\begin{figure}
\centering\includegraphics[scale=0.5]{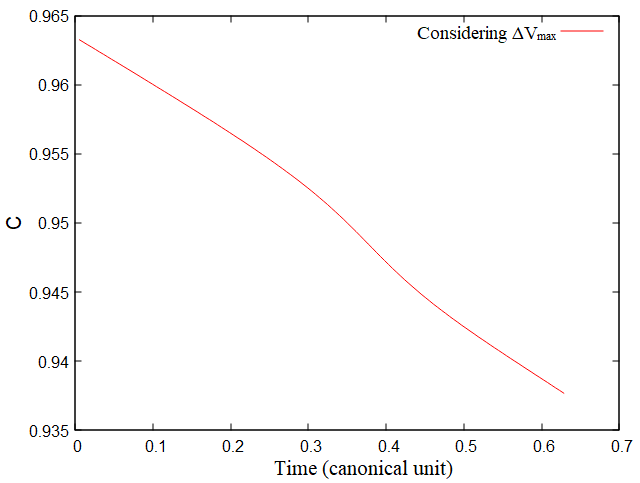}
\caption{Jacobi constant versus transfer time during the transfer from $L_2$ to $L_1$ with $t$ minimum.}
\label{cfast}
\end{figure}

Figure \ref{cfast} provides information on the Jacobi constant when we consider the shortest transfer time t = 0.1 day. As in the previous case, Jacobi's ``constant'' decreases over time. Note that the value of C is smaller compared to the previous case, which means that the spacecraft has more energy. This has already been mentioned before, as more velocity (thus more energy) is needed to reach equilibrium in less time.

\section{Conclusions}
\label{conclusion}

The exact locations of the five Lagrange points in a recently discovered binary asteroid were determined, as well as the energy (Jacobi constant) associated with those equilibrium points. All Lagrange points are in the plane of the primaries. Lagrange points are regions in space where there is an equilibrium of forces considered in the model. In the presence of other forces, they are no longer equilibrium points, but still have a small resultant force compared to other points, so they are great locations for maintaining a spacecraft with minimum station-keeping maneuvers.

The PSS is a powerful tool that makes it possible to find periodic orbits around a binary system. We built these maps considering the CRTBP in a rotating system. By restricting the energy level (or Jacobi constant), we reduce our system by one dimension so, considering the planar problem, the phase space is reduced to three dimensions and, consequently, the qualitative dynamical behavior of trajectories may be investigated by their intersections with a suitable Poincar\'e section.
The energy levels used are the ones of the equilibrium points. 

We found stable periodic orbits (prograde and retrograde), around each primary, with periods 1, 6, 8, 11, depending on the Jacobi constant used when we do not take into account disturbances from other celestial bodies as well as the size of the primaries.

We built tables that provide the initial conditions that generate the periodic orbits for the 2017 YE5 system. These tables also inform the stability of the periodic orbits found.

We then numerically investigate the influence of SRP on a particle in the vicinity of the 2017 YE5 system. We have seen that periodic orbits no longer exist, but it is still possible to find orbits that survive through all numerical integration (6 months) around the primary bodies.

We found several regions where the forward and retrograde orbits of a spacecraft survive over the integration time (six months) when solar radiation pressure is taken into account. Numerical evidence shows that retrograde orbits have larger regions of initial conditions that generate orbits that survive for six months compared to direct orbits.

Finally, we calculate the minimum values that an impulsive maneuver needs to allow transfers between collinear equilibrium points. We showed that, during the time of transfers, the influence of the Sun does not significantly alter the energy needed to perform the maneuver nor the trajectory taken by the space vehicle.

\section*{Acknowledgements}

The authors wish to express their appreciation for the support provided by: grants 150678/2019-3 and 309089/2021-2 from the National Council for Scientific and Technological Development (CNPq); grants 2016/18418-0 and 2016/24561-0 from S\~ao Paulo Research Foundation (FAPESP); grants E-26/201.877/2020 from the Foundation for Research Support of the State of Rio de Janeiro (FAPERJ); grant 88887.374148/2019-00 from the Coordination for the Improvement of Higher Education Personnel (CAPES) and grants 0000-000 from RUDN University. From Nacional Observatory (ON) and from the National Institute for Space Research (INPE). This publication has been supported by the RUDN University Scientific Projects Grant System, project No 202235-2-000. This work is supported by the European Regional Development Fund (FEDER), through the Competitiveness and Internationalization Operational Programme (COMPETE 2020) of the Portugal 2020 framework [Project SmartGlow with Nr. 069733 (POCI-01-0247-FEDER-069733). This work is funded by FCT/MCTES through national funds and when applicable co-funded EU funds under the project UIDB/50008/2020-UIDP/50008/2020. Lima, N.B. also thanks L'Or\'eal-UNESCO-ABC ``For Women in Science''. NBDL, EH, and TFAS thank CNPq for the scientific productivity scholarship. This research also was supported by Brazilian Agencies FACEPE and FINEP.

\bibliographystyle{elsarticle-harv}







\end{document}